\documentstyle[12pt,epsf,aaspp4]{article}

\newcommand{\etal}{et al.\ }

\newcommand{\lya}{Ly$\alpha$ }

\begin{document}

\title{A Fast, Accurate and Robust Algorithm For Transferring Radiation in Three-Dimensional Space}
\author{Renyue Cen\altaffilmark{1}}

\altaffiltext{1} {Princeton University Observatory, Princeton University, Princeton, NJ 08544; cen@astro.princeton.edu}

\begin{abstract} 
We have developed an algorithm for transferring 
radiation in three-dimensional space.
The algorithm computes radiation source and sink terms 
using the Fast Fourier Transform (FFT) method,
based on a formulation in which 
the integral of any quantity (such as emissivity or opacity)
over any volume may be written in the classic convolution form.
The algorithm is fast with
the computational time 
scaling as $N(\log N)^2$, where $N$ is the number of grid points
of a simulation box,
independent of the number of radiation sources.
Furthermore, in this formulation
one can naturally account for both local radiation sources
and diffuse background as well as any extra external sources,
all in a self-consistent fashion.
Finally, the algorithm is completely stable and robust.

While the algorithm is generally applicable,
we test it on a set of problems that encompass
a wide range of situations in cosmological applications,
demonstrating that the algorithm is accurate.
These tests show that the algorithm 
produces results that are in excellent agreement with analytic
expectations in all cases.
In particular, 
radiation flux is guaranteed to propagate in the right direction,
with the ionization fronts traveling at the correct speed with an
error no larger than one cell for all the cases tested.
The total number of photons is conserved in the worst
case at $\sim 10\%$ level and typically at $1-5\%$ level 
over hundreds of time steps.
As an added advantage, 
the accuracy of the results depends weakly on the size
of the time step, with a typical cosmological
hydrodynamic time step being sufficient.

\end{abstract}


\section{Introduction} 
Radiative transfer in three dimensional space
is a seven-dimensional problem:
three spatial dimensions, 
two angular dimensions, one frequency dimension plus
the time dimension.
As a result, although the basic physics involved is well understood,
a direct computation of radiative transfer in
three dimensional space is prohibitively costly.

However, radiation field is known to 
play a very important role in determining
the ionizational and thermodynamic state of the cosmic gas.
Rudimentary treatments of radiative transfer
from assuming a uniform radiation field  
(e.g., Cen \& Ostriker 1993) 
to using the somewhat improved method with local optical depth
approximation (Gnedin \& Ostriker 1997; Cen \& Ostriker 1999) 
have afforded the breakthrough computations of the optically thin regions
of the \lya forest 
with notable successes
(Cen \etal 1994; Zhang \etal 1995; Miralda-Escud\'e \etal 1996;
Hernquist \etal 1996).
Nevertheless, 
fluctuations in the ionizing radiation field 
are expected to exist even in the optically thin regions, 
since the radiation sources 
(galaxies and quasars) as well as density fields
are known to be significantly clustered at all times.
Turning away from optical thin regions,
one is faced with 
transitional regions of optical depth of order unity
and optical thick regions 
largely shielded from external radiation.
Moreover, radiation sources themselves most likely
reside in dense, at least partially shielded regions
often with complicated structures and geometries. 
For all these regions a proper treatment of radiative transfer
is clearly demanded.
The predictive power of any theory of galaxy and structure formation,
especially
with respect to the cosmological reionization, the formation and 
evolution of \lya forest, damped \lya systems and galaxies/stars,
is fundamentally limited until one can accurately
treat this essential process of radiative transfer.

Significant progress has been made in recent years
in developing practical
algorithms for radiative transfer in cosmological
applications.
Most of the effort has largely been concentrated
on two types of approaches:
the ray tracing approach (Abel \etal 1999; Rozoumov \& Scott 1999;
Kessel-Deynet \& Burkert 2000)
and moment equations 
approach (Stone, Mihalas \& Norman 1992 for two dimensional
space; Norman, Paschos, \& Abel 1998; Gnedin \& Abel 2001).
Several Monte Carlo methods have also been explored
(Sokasian \etal 2001; Ciardi \etal 2001).
The primary limitation of the ray tracing algorithm
is its intrinsic high cost to 
follow a large number of radiation sources.
The moment equations approach condenses the intrinsic
difficulty of treating long range radiation propagation
out as the Eddington tensor.
The problem then boils down to evaluating 
the Eddington tensor accurately and efficiently.
The recent suggestion by Gnedin \& Abel (2001)
of computing the Eddington tensor by ignoring optical depth 
is novel and clearly worth being explored further.

In this paper we develop an entirely new algorithm
for transferring radiation in three dimensional space.
In essence, this algorithm computes the optical depth 
between any pair of points
and the source function for each point
using convolution techniques. 
The angular discretization is performed at the 
receiving site, rather than at the source site as in the normal ray tracing
scheme,
which guarantees coverage of all space regardless of the fineness
of the angular discretization.
The overall scaling of the cost with this algorithm is $N(\log N)^2$.
Unlike conventional ray casting schemes,
the computational cost with the present algorithm 
is independent of the number of sources,
suitable for problems encountered in cosmological simulations
where a large number of direct, ionizing sources (e.g., small galaxies)
as well as secondary, processed ionizing sources (e.g., scattered 
photons, recombination photons) are present.
This paper is organized as follows.
In \S 2 we describe the basic algorithm.
In \S 3 we present a battery of tests to show that
the present algorithm is accurate.
Discussion and conclusions are 
given in \S 4.

\section{A New Algorithm for Radiative Transfer in Three-Dimensional Space} 

In the rest frame of a bundle of photons
the equation of radiation transfer for those photons (Spitzer 1980) is:

\begin{equation}
{d I_{\nu}\over ds} = -\kappa_\nu I_\nu + j_\nu,
\end{equation}

\noindent 
where $s$ is the path length
and $I_{\nu}(\vec x, \vec n, t)$ is the specific intensity 
with $I_\nu d\nu d\omega dA dt$ being 
the energy during a time interval $dt$ passing 
through an area 
$dA$
about a spatial point $\vec x$,
within a frequency interval $d\nu$,
within the solid angle $d\omega$ about $\vec n$;
$\kappa_\nu(\vec x,t)$
and $j_{\nu}(\vec x,t)$
are the opacity and emissivity at $\vec x$ at time $t$.
We can integrate equation (1) and write it in an integral form
for the specific flux $F_{\nu}$ defined as

\begin{equation}
F_{\nu}(\vec x,t) \equiv \int I_\nu(\vec x, \vec n,t) d\omega
\end{equation}

\noindent 
at position $\vec x$ and time $t$:

\begin{equation}
F_{\nu}(\vec x,t) = {1\over 4\pi}\int\int j_\nu(\vec x+s\vec n,t) e^{-\tau_\nu(\vec x, s\vec n,t)} ds d\omega,
\end{equation}

\noindent 
where 
$j(\vec x + s\vec n, t)$ 
is the emissivity at distance $s$ from position $\vec x$
in the direction $\vec n$ (an unit vector) at time $t$;
the two integrals are over path length and solid angle, respectively;
$\tau_\nu(\vec x, s\vec n, t)$ is the optical depth from 
position $\vec x$ to position $\vec x + s\vec n$ at time $t$:

\begin{equation}
\tau_\nu(\vec x, s\vec n, t) = \int_0^r \kappa_\nu(\vec x, s^\prime\vec n,t) ds^\prime,
\end{equation}

\noindent 
where $\kappa(\vec x, s\vec n, t)$ 
is the opacity at distance $s$ from position $\vec x$
in the direction $\vec n$ at time $t$.

The implicit principal assumption that has been made to 
derive Equation (3) is that 
the speed of light is infinity,
analogous to the conventional treatment of gravitational interactions. 
Although this assumption may not be necessary,
it makes implementation of this method especially simple and
we adopt it.
This assumption is generally excellent for cosmological simulations,
where light crossing time over a simulation box 
is indeed much smaller than the typical hydrodynamic time step.
It should, however, be noted that this assumption
does not impose any practical restrictions on the propagation speed
of ionization fronts.

Equation (3) may be discretized in the following form:

\begin{equation}
F_{\nu}(\vec x,t) =  \sum_{s} \sum_{\vec n} S_\nu(\vec x,s\vec n,t)  e^{-\tau_\nu(\vec x, s\vec n,t)} \left({1\over 4\pi s^2}\right),
\end{equation}

\noindent 
where the two sums are over the ray path (assuming to be
a straight line) and the solid angle, respectively,
and the source term $S_\nu$ is defined as

\begin{equation}
S_\nu(\vec x,s\vec n,t)\equiv {\bar j_\nu(\vec x,s\vec n,t)\Delta V(\vec x, s\vec n}) 
\end{equation}

\noindent 
with $\Delta V(\vec x, s\vec n)$ 
being the discretization volume element about position $\vec x + s\vec n$
and $\bar j_\nu$ is the mean emissivity over $\Delta V(\vec x, s\vec n)$.
The problem now translates to the evaluation of 
the two terms on the right hand side in Equation (5),
$S_\nu(\vec x, s\vec n,t)$ and $e^{-\tau_\nu(\vec x, s\vec n,t)}$.
This is the core of the overall computation
and how it is computed
determines the effectiveness and accuracy of the algorithm.
We propose to evaluate these two terms in a novel way.

To proceed further 
we will now choose the spherical coordinate system
for the radiation field about $\vec x$,
under which we will discretize the angular and radial dimensions
in a spherically symmetric fashion.
This is to say, 
one can always find a pair of equal volume elements (of arbitrary domain shape)
one about point $\vec x$ and the other point $-\vec x$ related by

\begin{equation}
\Delta V(\vec x, r\vec n) = \Delta V(\vec x, -r\vec n),
\end{equation}

\noindent 
where $r$ is the radial distance from $\vec x$ along a radial
direction $\vec n$ or $-\vec n$.
For every vector $y=r\vec n$ that lies in 
$\Delta V(\vec x, r\vec n)$, one can find an opposite vector (about $\vec x$)
$-\vec y= r(-\vec n)$ that lies in 
$\Delta V(\vec x, -r\vec n)$.
This property will be essential in our implementation of 
the computation of the two terms in Equation (5).
Let us rewrite the source term $S_\nu$ in Equation (6) as

\begin{equation}
S_\nu(\vec x,r\vec n,t) = \int_{\Delta V} j_\nu(\vec x+r\vec n,t) dV,
\end{equation}

\noindent
where the integration domain is the volume element 
$\Delta V(\vec x, r\vec n)\equiv r^2 drd\omega(r,\vec n)$
about position $\vec x + r \vec n$. 
It is now time to make a simple but algorithmically essential
modification of Equation (8) by inserting
a window function in the integral:

\begin{equation}
S_\nu(\vec x,r\vec n,t) = \int j_\nu(\vec x + r\vec n,t) g(\vec r) d^3\vec r,
\end{equation}

\noindent
where $g(\vec r)$ is a window function 
about the origin $\vec x$      
with the following property:
\begin{eqnarray}
g(\vec r) &=& 1 {\ \ \ \rm for\ vectors\ within\ }\Delta V\hfill \nonumber\\
          &=& 0 {\ \ \ \rm otherwise.}\hfill
\end{eqnarray}
\noindent
Note that the integral in Equation (9) is now over the entire space
about $\vec x$.
The reader is then asked to 
make the following critical observation:
for each window function $g(\vec r)$ there is another window function
$f(\vec r)$,
which is spherically symmetrically paired with it,
i.e., 
\begin{equation}
f(-\vec r) = g(\vec r),
\end{equation}

\noindent
as indicated by Equation (7).
Replacing $g(\vec r)$ with $f(-\vec r)$ in Equation (9) gives,

\begin{equation}
S_\nu(\vec x,r\vec n,t) = \int j_\nu(\vec x + r\vec n,t) f(-\vec r) d^3\vec r.
\end{equation}

\noindent
Changing integration variable from $\vec r$
to $\vec y\equiv \vec x+ \vec r$ Equation (12) becomes

\begin{equation}
S_\nu(\vec x,r\vec n,t) = \int j_\nu(\vec y,t) f(\vec x - \vec y) d^3\vec y.
\end{equation}

\noindent
The integration domain is still over the entire space
but the window function $f(\vec x -\vec y)$ 
now has a moving origin $\vec x$.
Equation (13) is of the standard convolution form and
can be efficiently evaluated for all points $\vec x$
simultaneously using the Fast Fourier Transform (FFT) techniques,
as the Fourier transforms are related by

\begin{equation}
\underline S_{\vec k} = \underline j_{\vec k} \underline f_{\vec k}, 
\end{equation}

\noindent
where $\underline S_{\vec k}$, $\underline j_{\vec k}$ and 
$\underline f_{\vec k}$ 
are Fourier transforms of $S_\nu$, $j_\nu$ and $f$, respectively.

The same formulation can be constructed for 
the optical depth $\tau_\nu (\vec x, r\vec n,t)$.
Here we instead compute the mean opacity 
of each volume element 

\begin{eqnarray}
\bar\kappa_\nu (\vec x,r\vec n,t) &=& {\int_{\Delta V} \kappa_\nu(\vec x + r\vec n,t) dV \over \Delta V} \nonumber\\
&=& {\int \kappa_\nu(\vec y,t) f(\vec x-\vec y) d^3\vec y \over \Delta V}
\end{eqnarray}

\noindent
using the same technique.
Then we integrate $\bar\kappa_\nu$ radially
along $\vec n$ in real space to obtain $\tau_\nu (\vec x, r\vec n,t)$.

\begin{equation}
\tau_\nu (\vec x, r\vec n,t) = \int_0^r\bar\kappa_\nu (\vec x, r^\prime\vec n,t) dr^\prime.
\end{equation}

\noindent
Using the FFT method,
$S_\nu(\vec x,r\vec n,t)$ (Equation 13) and $\bar\kappa_\nu(\vec x,r\vec n,t)$
(Equation 15)
for all grid points of a simulation box with respect to all other
points in the simulation box can be evaluated
simultaneously with the total cost scaling
as $N(\log N) m_r m_a$, where $N$ is the total number of grid points,
$m_r$ is the number of radial bins along each cone
and $m_a$ is the number of angular discretization elements.

The choice of $m_r$ and how to discretize the radial direction
may be problem dependent. 
It is, however, illuminating to note that
both radiation flux and gravitational force
follow the same inverse square law.
It is also noted that
the effect due to optical depth attenuation tends to relatively
suppress the contribution from distant sources.
Thus, one may conservatively apply some of the techniques and ideas
used in gravitational tree codes (Barnes \& Hut 1986).
Specifically, one can more coarsely sample more distant regions
to discretize the radial direction,
the simplest of which would result in $m_r\propto \log N$.
The choice of $m_a$ requires some experiments. 
We note that, in the present algorithm, any coarse level of 
angular discretization guarantees that all sources inside the simulation
box are covered.
This property is in contrast with the conventional ray casting
scheme, where a large number of rays from a source
is necessary in order to intersect every cell in the simulation box at 
least once.
The primary operational difference is that in the present algorithm
the operation is ``gathering" flux 
by the receiving region under consideration, 
whereas in the ray casting method the operation is ``broadcasting" flux
to each receiving region.
Testing shows that $m_a=16\times 16=256$ to cover
the full solid angle $4\pi$ appears adequate.
Combining the three scaling factors 
gives the total number of operations
for evaluating $S_\nu$ or $\tau_\nu$
for all grid points  at each time step:

\begin{equation}
{\rm Total\ number\ of\ operations\ } N_{op} = A_{\rm FFT} N(\log N)^2 m_a,
\end{equation}

\noindent
where $A_{\rm FFT}$ is the normal 
prefactor of the FFT operation.
Note that the algorithm presented here 
has the desired property that its computational cost
is independent of the number of radiation sources.
As a matter of fact, each grid point is treated as both a
source and sink. This property allows us to compute 
problems where a large number of radiation sources 
are present. Examples include reionization of the universe
by sub-galactic galaxies and radiation field
where  
diffuse sources such as recombination photons and scattered photons
contribute significantly.

We will now implement the algorithm  
in a three-dimensional
simulation box with a uniform mesh for hydrodynamic variables
and periodic boundary conditions for both hydro quantities and 
radiation field.
Two separate coordinate systems and two correspondingly
separate grids are in use now:
the Cartesian coordinate system
for the hydrodynamic quantities with a uniform Cartesian grid
and the spherical coordinate system
for the radiation field about
each hydro grid point with a spherical grid.
The angular discretization about each 
hydro grid point is made uniformly across the solid
angle $4\pi$ with $m_a$ elements each having a full solid angle
of $4\pi/m_a$.
The radial discretization is performed using $m_r$ bins
from $r=0$ to $r_{max}=\sqrt{3}L$ spaced uniformly
in logarithm maximizing the shape of each volume element
as a cube, where $L$ is the simulation box size.
The $m_r m_a$ window functions (i.e., $f$ in Equations 13,15)
for each discretized element of the radiation grid in the spherical
coordinates are computed once initially and their Fourier
transforms are stored and used for all subsequent time steps.

So far we have operated in a static coordinate system.
In cosmological applications one needs to take into account
the cosmological effects. 
We will separate the overall radiation field into two parts:
1) radiation flux due to sources within the simulation box,
and
2) external radiation flux.
As long as the simulation box is much smaller
than the Hubble radius,
this division allows us to split these two kinds of sources
cleanly in the simple fashion:
radiation flux from (1) is treated as local,
which is not affected by cosmic expansion,
and radiation flux from (2) is 
diffuse and subject to the cosmological effects.
The two kinds of radiation fields are related and computed
in a self-consistent way.
We keep track of radiation that leaves the simulation box
and add it to the diffuse background (see below)
and every point in the simulation box 
is subject to the diffuse background 
with self-consistently computed attenuation.
Thus, the total flux at any point inside the simulation box is 

\begin{eqnarray}
F_{\nu,tot}(\vec x,t) = F_{\nu,in}(\vec x,t) + F_{\nu,ext}(\vec x,t),
\end{eqnarray}

\noindent
where 
$F_{\nu,in}(\vec x,t)$ is the contribution from sources
inside the simulation box in Equation (3) and 
$F_{\nu,ext}(\vec x,t)$ is the contribution from the
diffuse background related to the specific intensity of the 
diffuse background at time $t$, $I_{\nu,diff}(t)$:

\begin{eqnarray}
F_{\nu,ext}(\vec x,t) &=& I_{\nu,diff}(t) \int e^{-\tau_\nu(\vec x, r_{upp}\vec n,t)}d \omega (\vec n) \nonumber \\
&=& I_{\nu,diff}(t) {4\pi \over m_a}\sum_{l=1}^{m_a} e^{-\tau_{\nu,l}(\vec x, r_{upp}\vec n,t)},
\end{eqnarray}

\noindent
where the integral or the sum 
is over the solid angle about position $\vec x$;
$\tau_\nu(\vec x, r_{upp}\vec n, t)$ is total optical depth 
along the cone about $\vec n$ from position $\vec x$ 
to a distance of $r_{upp}$
at time $t$ ($r_{upp}$ is the upper radius of that cone
given the constraint that the simulation box is periodic;
$r_{upp}$ is orientation-dependent 
because the hydro simulation box is not spherical 
about each grid point for the radiation field;
the range for $r_{upp}$ is $L \le r_{upp} \le r_{max}=\sqrt{3}L$).
The evolution of the specific intensity of the 
diffuse background, $I_{\nu,diff}(t)$,
is governed (Cen \& Ostriker 1993) by

\begin{equation}
{\partial I_{\nu,diff}(t)\over \partial t} = H(t)\left[\nu{\partial I_{\nu,diff}(t)\over \partial \nu}-3I_{\nu,diff}(t)\right] + c s_{\nu,in}(t) + c s_{\nu,ext}(t) - c\bar\kappa_{\nu,diff}(t) I_{\nu,diff}(t),
\end{equation}

\noindent
where $H(t)$ is the Hubble constant, $c$ is the speed of light
and the mean opacity $\bar\kappa_{\nu, diff}(t)$ 
at time $t$ is determined self-consistently by 
averaging the opacities in the simulation box properly
taking into account optical shielding effect:

\begin{eqnarray} 
\bar\kappa_{\nu, diff}(t) &=& {1\over 4\pi N}\sum_{\vec x} \kappa_{\nu}(\vec x)\int e^{-\tau_\nu(\vec x, r_{upp}\vec n,t)}d\omega(\vec n) \nonumber \\
&=& {1\over N m_a}\sum_{\vec x} \kappa_{\nu}(\vec x)\sum_{l=1}^{m_a} e^{-\tau_{\nu,l}(\vec x, r_{upp}\vec n,t)},
\end{eqnarray}

\noindent
where the outer sum is over all $N$ grid points of the simulation box each
having an opacity $\kappa_{\nu}(\vec x)$ and 
inner sum is over the $m_a$ solid angles about each point.
The term on the right hand side of Equation (20) due to 
sources inside the simulation box is
$s_{\nu,in}$, which may be computed from emissivity 
$j_{\nu}(\vec x)$ of each hydro grid point
of the simulation box:

\begin{eqnarray} 
s_{\nu,in}(t) &=& {1\over 4\pi N}\sum_{\vec x} j_{\nu}(\vec x)\int e^{-\tau_\nu(\vec x, r_{upp}\vec n,t)}d\omega(\vec n)  \nonumber \\
&=& {1\over Nm_a}\sum_{\vec x} j_{\nu}(\vec x)\sum_{l=1}^{m_a} e^{-\tau_{\nu,l}(\vec x, r_{upp}\vec n,t)};
\end{eqnarray}

\noindent
this is the radiation due to internal sources that leaves the simulation box.
There may be additional sources to the diffuse background.
For example, one may wish to take into account rare sources
such as quasars that are outside the simulation box.
This can be achieved by simply adding the needed terms as
$s_{\nu,ext}(t)$.

The formulation of Equation (15) deserves more discussion.
The integration over the 
domain $\Delta V$ is effectively collapsing
the angular dimensions to lump
together absorbing material into a radial ``line"
of length $\Delta r$ for that volume element.
This could cause an overestimate of opacity for radiation
sources that are located inside the same volume 
element or behind the considered volume element
but are spatially displaced perpendicular to the line of sight
with respect to the absorbing material.
In a more extreme example,
a handful of cells of very high optical depth 
may totally block out radiation downstream,
if the mean optical depth is evaluated
by linearly averaging optical depth spatially.
We propose to modify the original formulation (Equation 15)
as follows.

\begin{eqnarray} 
<e^{-\kappa_\nu (\vec x + s\vec n,t)\Delta L}> &=& {\int_{\Delta V} e^{-\kappa_\nu(\vec x + s\vec n,t) \Delta L} dV \over \Delta V} \nonumber \\
&=& {\int e^{-\kappa_\nu(\vec y,t)\Delta L} f(\vec x-\vec y) d^3\vec y \over \Delta V}
\end{eqnarray} 

\noindent
and 
\begin{equation}
\bar\kappa_\nu (\vec x + s\vec n,t) = -{1\over \Delta L}\ln <e^{-\kappa_\nu (\vec x + s\vec n,t)\Delta L}>,
\end{equation}

\noindent
where $\Delta L$ is the size of a hydro cell.
This modified formulation is in some sense
like averaging the flux instead of the optical depth.
This physically motivated modification 
quite successfully circumvents the forementioned possible
problems in realistic situations, as subsequent tests will show (see \S 3). 
We adopt this particular scheme to present
quantitative results in the next section (\S 3).
We note that in the case of a uniform opacity Equation (15)
and Equations (23,24) would yield identical results.

Let us briefly summarize the salient features of the proposed
algorithm.
What we have shown is that the algorithm is {\it fast},
scaling as $N(\log N)^2$, where $N$ is the number of grid points
in a simulation box, independent of the number of radiation sources.
Given that computation does not involve any differentiation
and all computed integrals are guaranteed to converge in all situations,
the algorithm is completely {\it stable and robust}.
We emphasize that 
the proposed algorithm, by design,
guarantees that flux propagates in the {\it right direction}
in any situation.
The tests presented in the next section 
will show that the amplitude of flux is also computed
with very small errors, indicating that 
the algorithm is also {\it accurate}.

\section{Tests of the Algorithm} 

We will present a suite of tests,
which are intended to have significant bearings 
on cosmological applications.
We will show that the method performs 
very well in all tested situations.
Clearly, fine tuning of the scheme is worth exploring 
and is deferred for future work.

For the tests shown we use a $32^3$ Cartesian grid with 
periodic boundary conditions.
We use $m_a=256$ uniformly spaced angular bin and
$m_r=18$ logarithmically  spaced radial bins for all cases.
For the tests shown we do not include cosmological affects;
i.e., $H(t)$ is set to zero.
But the accuracy of the algorithm should not be affected
by this simplification.


\subsection{Static Spherical Optical Depth Distribution}

\begin{figure}
\plotone{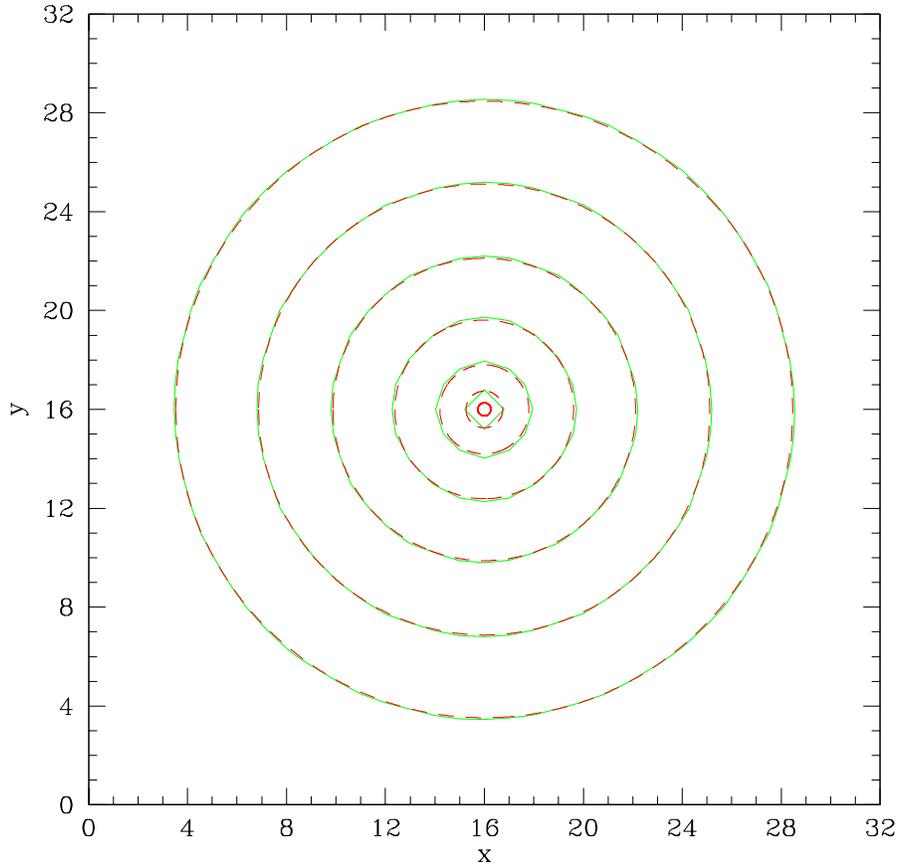}
\caption{
Distribution of flux in the x-y plane (with $z=16$)
for a source sitting at cell (16,16,16)
embedded in a static, uniform opacity distribution.
The contour levels are logarithmically spaced
with an increment of $1$ dex per contour level.
The dashed contours are the analytic results (Equation 25)
and solid contours are obtained with the present algorithm.
The units for the contours levels are arbitrary.
}
\label{Static}
\end{figure}

It is guaranteed that the proposed formalism will give the correct
flux distribution in the case of no optical attenuation.
The simplest non-trivial test then is a source embedded in a static,
uniform, non-zero opacity distribution.
The flux distribution about a single source is then

\begin{equation}
F(r) = {L\over 4\pi r^2} e^{-\bar\kappa r},
\end{equation}

\noindent
where $L$ is the luminosity of the source;
$r$ is source-centric radius;
$\bar\kappa$ is the opacity.
Figure \ref{Static} 
shows the flux distribution in the x-y plane with $z=16$, where the 
luminous source is located at cell $(16,16,16)$
and the opacity per cell is $\bar\kappa=0.5$.
The solid contours nearly overlap with the analytic
expectations (dashed contours) with some slight
deformations from sphericity simply due to the nature of the
Cartesian grid.
Clearly, the computed flux is accurate.

Since the equation for the flux is linear with respect
to the source luminosity,
the method would obviously yield 
results of the same accuracy for any number of distributed
sources.

\begin{figure}
\plotone{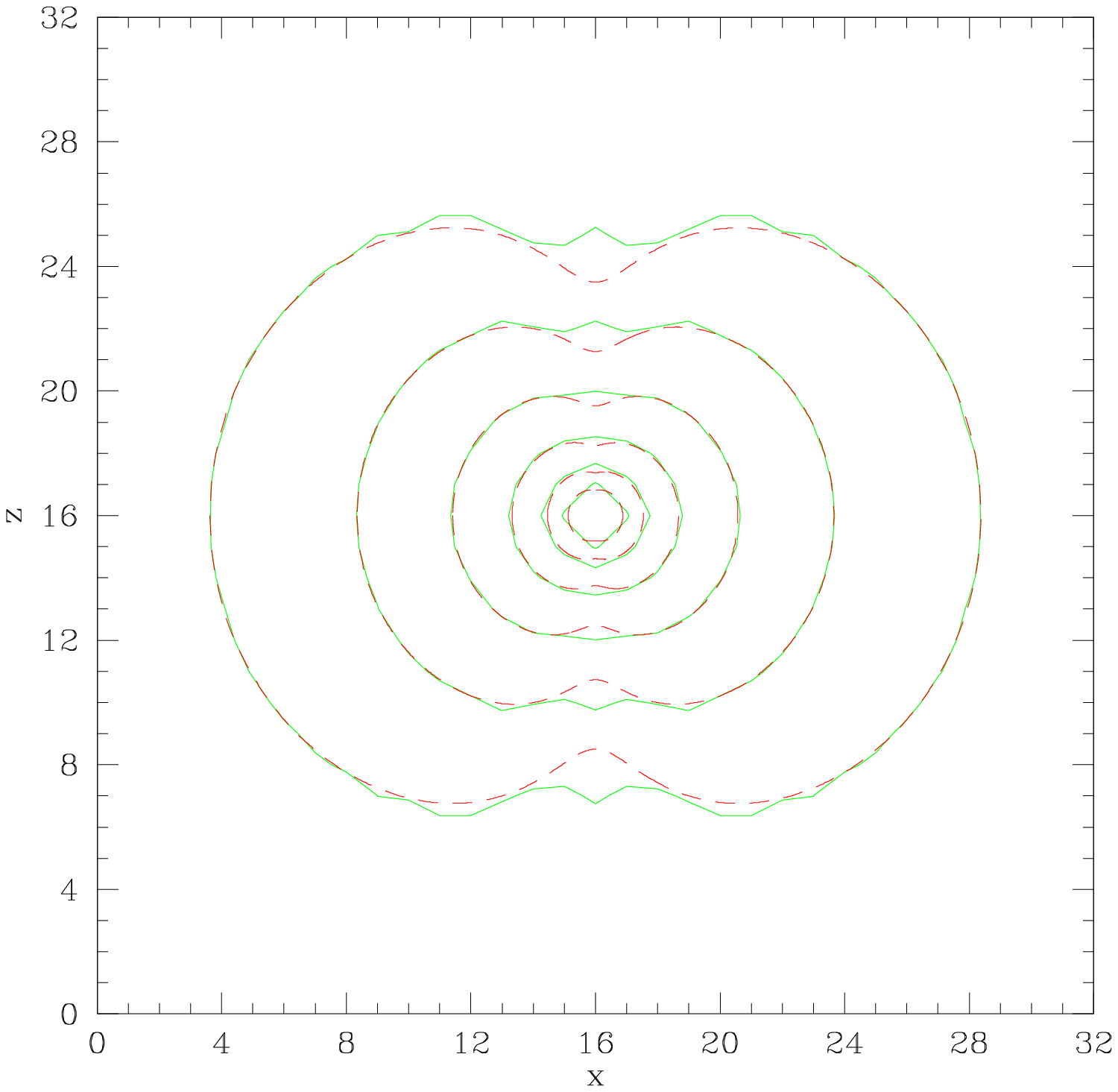}
\caption{
Distribution of flux in the x-z plane (with $y=16$)
for a source sitting at cell (16,16,16) 
embedded in a static, elliptical opacity distribution (Equation 26).
The contour levels are logarithmically spaced
with an increment of $0.5$ dex per contour level.
The dashed contours are the analytic results
and solid contours are obtained with the present algorithm.
The units for the contours levels are arbitrary.
}
\label{Ellip}
\end{figure}

\subsection{Static Elliptical Optical Depth Distribution}

The next step is to test  
a non-spherical static opacity distribution.
An axisymmetric elliptical distribution of opacity of the 
following form is tested:

\begin{equation}
\kappa(\theta) = {c b \over \sqrt{1-(1-b^2)\cos^2\theta}},
\end{equation}

\noindent
where $\theta$ is the angle with respect to the
positive $z$ direction, 
and $c$ and $b$ are the normalization and the
ellipticity parameter of the distribution, respectively;
$\kappa=c$ at $\cos\theta=1$ and 
$\kappa=cb$ at $\cos\theta=0$.

Figure \ref{Ellip} 
shows the computed flux distribution (solid contours)
in the x-z plane with $y=16$ 
compared to the analytic result (dashed contours).
We use $c=0.2$ and $b=0.2$ (see Equation 26) for this illustration.
We see that the results are quite satisfactory
given the Cartesian grid.
The agreement with the analytic result
is very good except near $\cos\theta=1$
where the opacity is the highest
and a maximum error of 2 cells is made there.
This is caused by a slight underestimate of the optical depth there,
due to the averaging scheme of optical depth
as indicated by Equations (23,24).
We have also tested similar cases by varying
either the opacity $c$ or ellipticity parameter $b$ (see Equation 26)
and find that the results have similar accuracies.

\subsection{Ionization of a Uniform Neutral Medium}

We now switch to a higher gear to test the algorithm
in a situation where the distribution of opacity changes self-consistently
with time due to ionization.
For this and subsequent tests we monitor photon 
number conservation by computing

\begin{equation}
\eta(t) = {N_e(t) + N_{diff}(t)\over N_{ph,tot}(t)},
\end{equation}

\noindent
where $N_e(t)$ is the number of free electrons created by time $t$
(assuming 100\% hydrogen)
in the simulation box
and $N_{diff}$ is the number of photons in the diffuse
background in the simulation box volume at time $t$ 
and $N_{ph,tot}$ is the total number of photons emitted by time $t$ 
by sources in the simulation box.
If photon number conservation is strictly observed,
$\eta$ would be unity.
In all the following tests
recombination time is set to infinity to make the problem
simple to track. But the accuracy of the method is unchanged
by this simplification.

\begin{figure}
\plotone{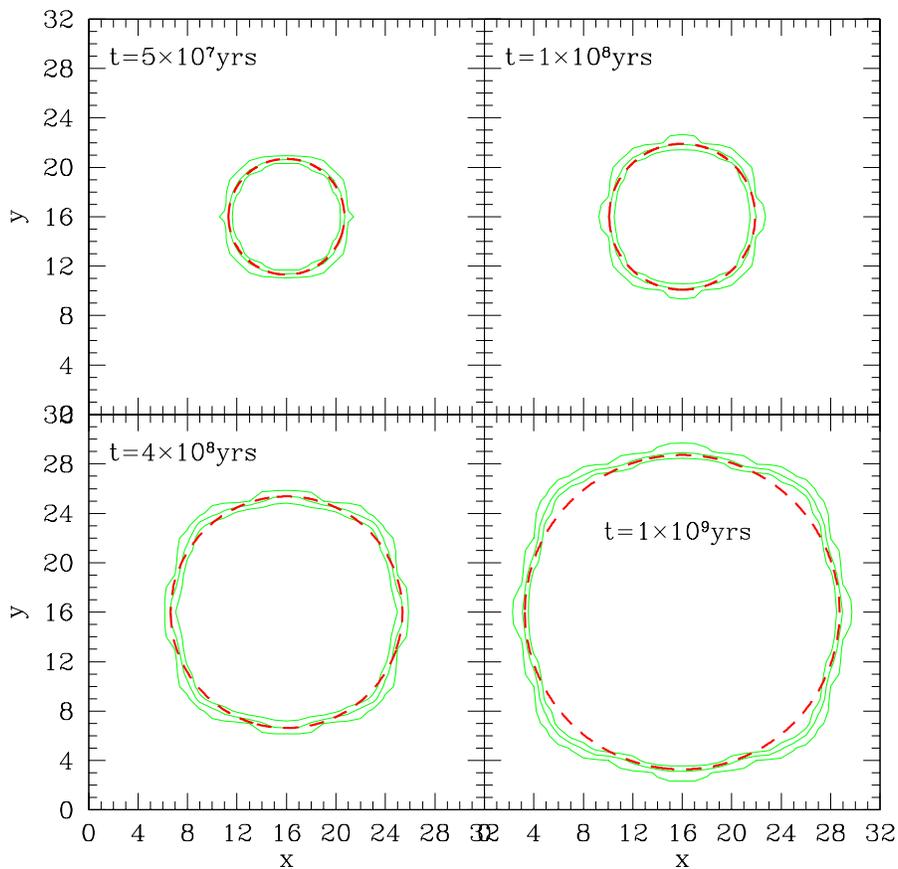}
\caption{
Distribution of neutral hydrogen fraction in the x-y plane (with $z=16$)
for a source of luminosity
$10^{51}~$photon/sec sitting at cell (16,16,16) 
in
a uniform density $n=1\times 10^{-3}~$cm$^{-3}$
at four epochs
$(5\times 10^7, 1\times 10^8, 4\times 10^8, 1\times 10^9)~$yrs.
The solid contours indicate the neutral hydrogen fraction
of $0.3, 0.6, 0.9$ inside out
computed with the present algorithm.
The dashed contours are the radii from analytic calculations (Equation 28).
}
\label{RHON0}
\end{figure}

\begin{figure}
\plotone{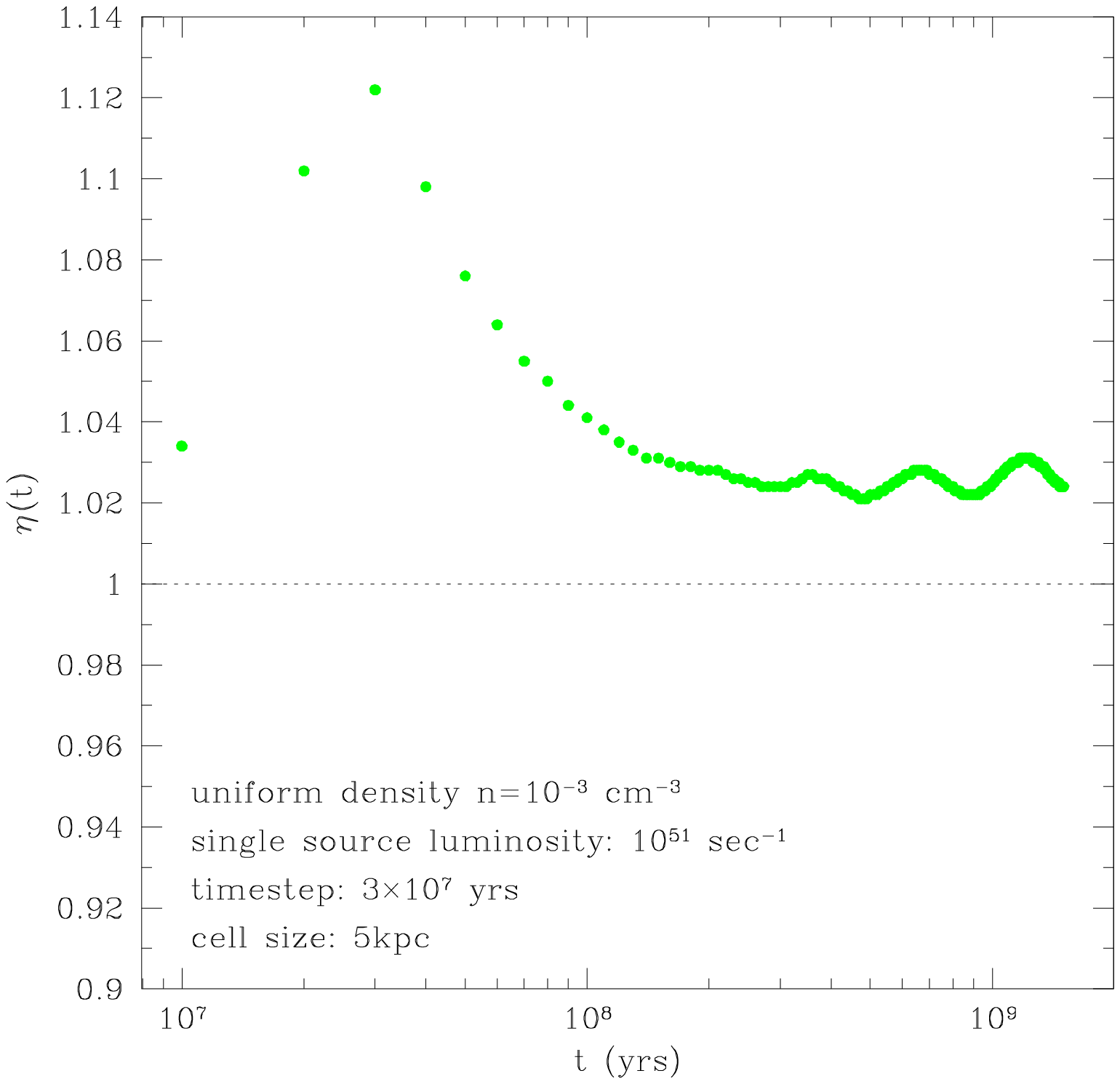}
\caption{
The error on photon number conservation (see Equation 27)
as a function of time
for the case of an ionizing galaxy
surrounded by a uniform density distribution 
(see Figure \ref{RHON0}).
}
\label{RHON0err}
\end{figure}

We start with the case of
a uniform neutral medium, consisting entirely of hydrogen atoms.
An ionizing source of ionizing luminosity $\dot N_{ph}$ photon/sec
is placed at cell $(16,16,16)$.
The ionizing photons from the source 
ionize the neutral medium outward and the radius of the ionization front,
the surface that separates the interior ionized medium
from the exterior neutral medium,
evolves with time approximately as:

\begin{eqnarray}
r(t) = 13.7 ({\dot N_{ph}\over 10^{51}})^{1/3} ({n\over 10^{-3} {\rm cm^{-3}}})^{-1/3} ({t\over 10^7 {\rm yrs}})^{1/3} {\rm kpc},
\end{eqnarray}

\noindent
where $n$ is the number density of the neutral hydrogen
and $t$ is the elapsed time.
The reason that this formula (26) is only approximate 
is that some photons may penetrate further out ahead
of the ionization front, if the optical depth per cell
is not sufficiently
high, and then the ``front" is no longer well defined;
the formula would be highly accurate for a high opacity
distribution. 
The optical depth per cell is
$\tau_\nu = \sigma_H n \Delta x = 15.4 (n/10^{-3} {\rm cm^{-3}}) (\Delta x/5 {\rm kpc})$,
where the hydrogen photo-ionization cross section
$\sigma_H=10^{-18}~$cm$^{2}$ is used throughout
and $\Delta x$ is the cell size.
Thus, for the present test, the optical depth per cell is very large
and formula (26) should be very accurate.

Figure \ref{RHON0} 
shows the contours of the neutral hydrogen fraction in the x-y plane
with $z=16$
at four epochs.
We use $\dot N_{ph}=10^{51}~$photon/sec, 
$n=10^{-3}~$cm$^{-3}$
and cell size of $\Delta x=5~$kpc.
The adopted gas density is about 15 times the mean gas density
at redshift $z=6$.
We see that the agreement between the computed results
and analytic expectations is excellent at all times.
The discrepancy on the radius
of the ionization front is evidently no larger than one cell.
Given the approximations that have been made in the implementation,
especially with respect to the computation of mean opacity 
(Equations 21,22), it is unclear how well the algorithm
conserves photons.
Figure \ref{RHON0err} shows the degree
of conservation of photons as a function of time.
It is seen that, except for the first 8 time steps,
the number of photons is conserved at better than
$4\%$ with the average at about $2-3\%$.
Timesteps of $10^7~$yrs are used but the results are insensitive 
to the timestep; using a timestep 3 times larger yield
very similar results except for the upper left corner panel
at the beginning of the simulation.
Note that the age of the universe at $z=6$ is about $5\times 10^8~$yrs.
The indicated convergence of results at this relatively
large size of timesteps is clearly advantageous.

\subsection{Ionization of an $r^{-1}$ Neutral Medium}

\begin{figure}
\plotone{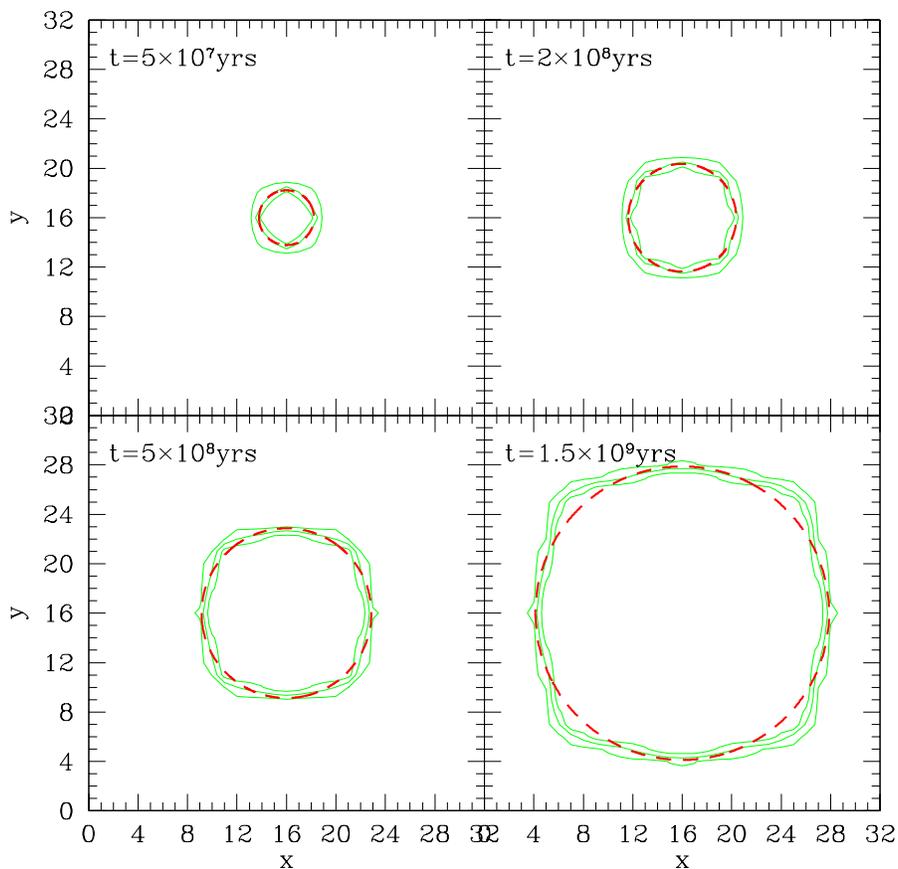}
\caption{
Distribution of neutral hydrogen fraction in the x-y plane (with $z=16$)
for a source sitting at cell (16,16,16)
in a density distribution of $n(r)\propto r^{-1}$ 
at four epochs
$(5\times 10^7, 2\times 10^8, 5\times 10^8, 1.5\times 10^9)~$yrs.
The dashed contours are the analytic results (Equation 29)
and solid contours are obtained with the present algorithm
indicating the neutral hydrogen fractions of $0.3, 0.6, 0.9$ inside out.
}
\label{RHON1}
\end{figure}

\begin{figure}
\plotone{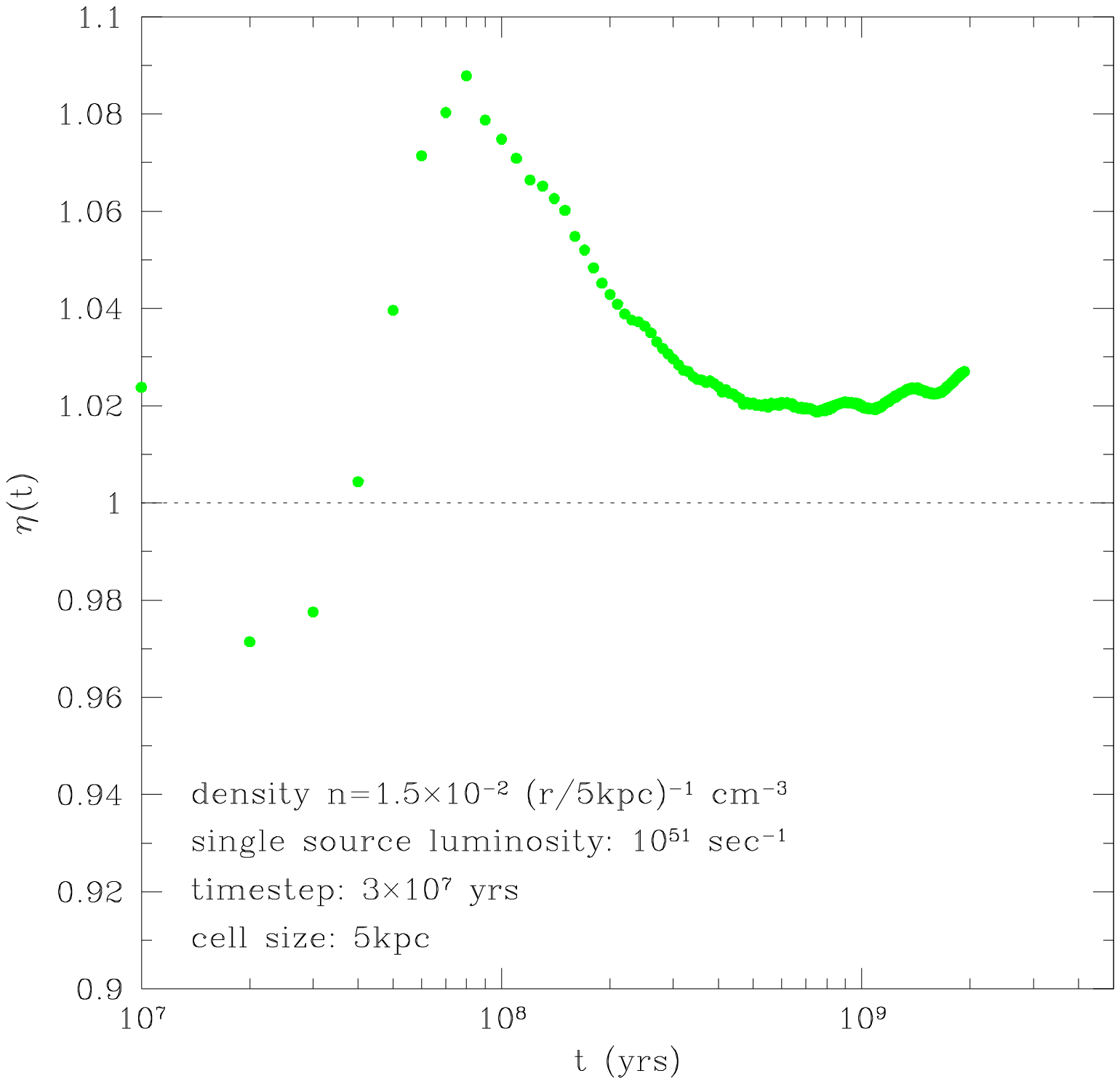}
\caption{
The error on photon number conservation as a function of time
for the case of an ionizing galaxy
surrounded by an $r^{-1}$ density distribution
(see Figure \ref{RHON1}).
}
\label{RHON1err}
\end{figure}

Next, we examine a case 
of a source sitting in a spherical density distribution,
$n = n_c (r/r_c)^{-1}~$cm$^{-3}$;
the central cell at $r=0$ is set to a density 
equal to $n_c (\Delta x/r_c)^{-2}$,
where $\Delta x$ is the cell size.
This density profile is close to that in the central
regions of dark matter halos (Navarro, Frenk, \& White 1997)
and is of significant
cosmological relevance.
The expected evolution of 
the radius of the ionization front is approximately:

\begin{eqnarray}
r(t) = \sqrt{35.2 ({\dot N_{ph}\over 10^{51}{\rm sec^{-1}}}) ({n_c\over 10^{-2} {\rm cm^{-3}}})^{-1} ({t\over 10^7 {\rm yrs}}) ({r_c\over 5 {\rm kpc}})^{-1} 
+ 25 \left(({3\over 4\pi})^{2/3} - {1\over 2\pi}\right)\left({\Delta x\over 5 {\rm kpc}}\right)^2} {\rm kpc},
\end{eqnarray}

\noindent
where the second (correction) term inside the
square root is due to the non-singular core of the modified
$r^{-1}$ density distribution.
The optical depth per cell as a function of source-centric radius is
$\tau(r) = 154 (n/10^{-2} {\rm cm^{-3}}) (\Delta x/5 {\rm kpc}) (r/r_c)^{-1}$.
Therefore, for the adopted values of $r_c$,
$n_c$ and $\Delta x$ (see below),
the formula will be very accurate over the range of radii examined
and the thickness of the ionization front should be
thin and close to one cell.

Figure \ref{RHON1} 
shows the contours of the neutral hydrogen fraction in the x-y plane
with $z=16$
at four epochs.
We use $\dot N_{ph}=10^{51}~$photon/sec, 
$n_c=1.5\times 10^{-2}~$cm$^{-3}$,
$r_c=5~$kpc 
and cell size of $\Delta x=5~$kpc.
Note that the adopted $n_c$ is approximately
$200$ times the mean gas density at $z=6$.
Timestep used is $10^7~$yrs but the results are insensitive 
to the timestep. 
From Figure \ref{RHON1} 
it is seen that the agreement between the computed results
and analytic expectations is excellent at all times
and the discrepancy on the radius
of the ionization front is no larger than one cell.
Figure \ref{RHON1err} presents the degree
of conservation of photons as a function of time,
showing that the number of photons is conserved at better than
$10\%$ with the average at about $2-3\%$ in this case.

\subsection{Ionization of an $r^{-2}$ Neutral Medium}

We turn to a steeper density distribution,
$n = n_c (r/r_c)^{-2}~$cm$^{-3}$;
the central cell at $r=0$ is set to a density 
equal to $n_c (\Delta x/r_c)^{-2}$,
where $\Delta x$ is the cell size.
The density profile is close to that of halos 
just interior of the virial radius 
(e.g., Tyson, Kochanski, \& dell'Antonio 1998).
The expected evolution of 
the radius of the ionization front is approximately:
\begin{equation}
r(t) = 3.4 ({\dot N_{ph}\over 10^{51}}) ({n_c\over 10^{-2} {\rm cm^{-3}}})^{-1} ({t\over 10^7 {\rm yrs}}) ({r_c\over 5 {\rm kpc}})^{-1} + 5 \left(({3\over 4\pi})^{1/3} - {1\over 4\pi}\right)({\Delta x\over 5 {\rm kpc}}){\rm\ kpc},
\end{equation}

\noindent
where the second term on the right hand side is due to 
the non-singular core imposed on the density profile.

\begin{figure}
\plotone{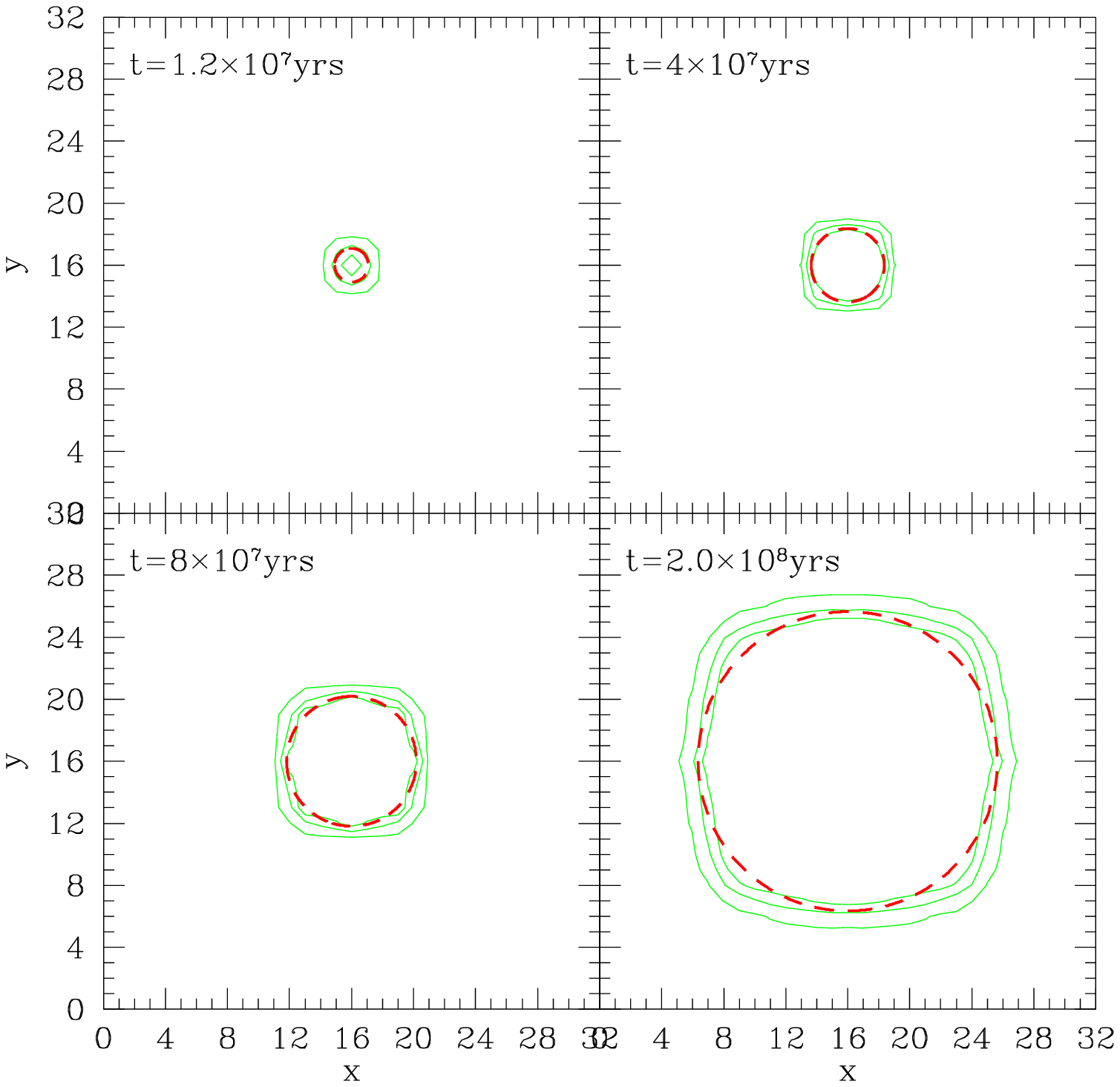}
\caption{
Distribution of neutral hydrogen fraction in the x-y plane (with $z=16$)
for a source sitting at cell (16,16,16)
in a density distribution of $n(r)\propto r^{-2}$
at four epochs
$(1.2\times 10^7, 4\times 10^7, 8\times 10^7, 2\times 10^8)~$yrs.
The dashed contours are the analytic results (Equation 30)
and solid contours are obtained with the present algorithm
indicating the neutral hydrogen fractions of $0.3, 0.6, 0.9$ inside out.
}
\label{RHON2}
\end{figure}

\begin{figure}
\plotone{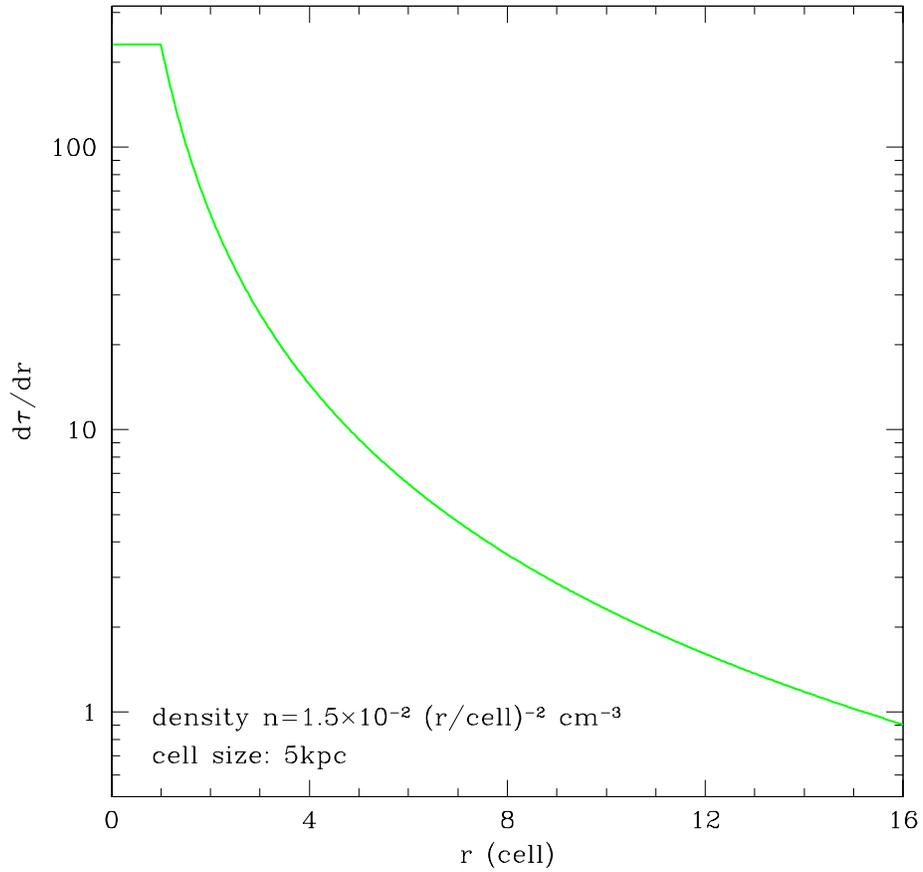}
\caption{
The original distribution of optical depth as a function of 
radius with an $r^{-2}$ density distribution.
We see that at $r\ge 6$ cells a
substantial amount of photons starts to
penetrate outward ahead of the ionization front
causing the slight thickening of the front,
see in Figure \ref{RHON2}.
}
\label{RHON2tau}
\end{figure}

\begin{figure}
\plotone{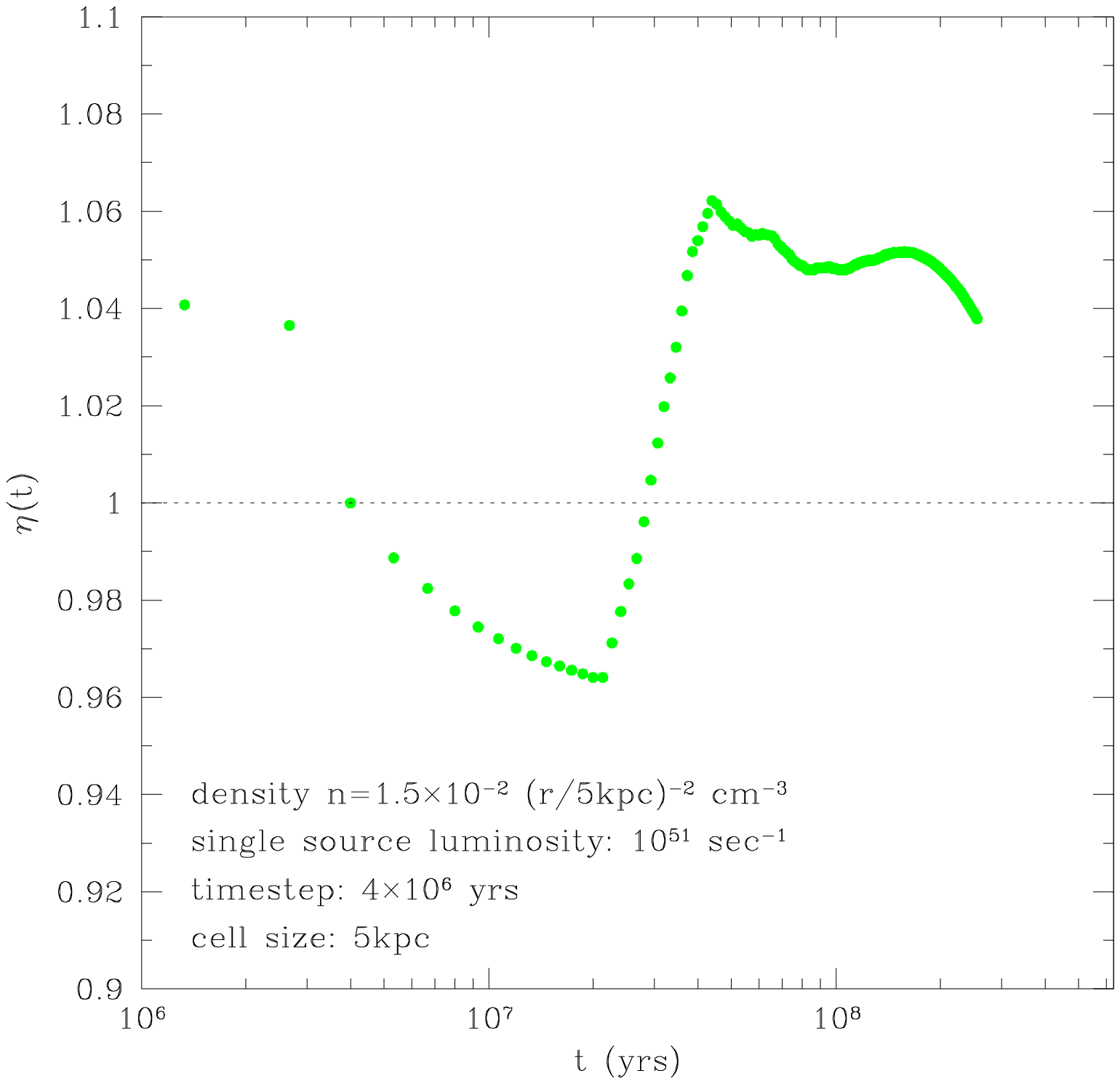}
\caption{
The error on photon number conservation as a function of time
for the case of an ionizing galaxy
surrounded by a $r^{-2}$ density distribution
(see Figure \ref{RHON2}).
}
\label{RHON2err}
\end{figure}

Figure \ref{RHON2} 
shows the contours of the neutral hydrogen fraction in the x-y plane
with $z=16$.
We use $\dot N_{ph}=10^{51}~$photon/sec, 
$n_c=1.5\times 10^{-2}~$cm$^{-3}$,
$r_c=5~$kpc 
and cell size of $\Delta x=5~$kpc
at four epochs.
Timestep used is $10^7~$yrs but like in other cases
the results are insensitive to the timestep.
The computed results
and analytic expectations agree very well at all times.
For the adopted values of $n_c$, $r_c$ and $\Delta x$,
the optical depth per cell as a function of source-centric radius is
$\tau(r) = 154 (n/10^{-2} {\rm cm^{-3}}) (\Delta x/5 {\rm kpc}) (r/r_c)^{-2}$
shown in Figure \ref{RHON2tau};
we have $\tau(r)=3.6$ for $r=8~$cells.
Thus, the computed shell of the ionization front
will be somewhat broadened at late times
due to a deeper penetration of ionizing photons outward of the
ionization front,
clearly evident in Figure \ref{RHON2}.
Figure \ref{RHON2err} shows
that the number of photons is conserved at better than
$6\%$ with the average at about $1-4\%$ for this case.

\subsection{Ionization of an Elliptical $r^{-2}$ Neutral Medium}

\begin{figure}
\plotone{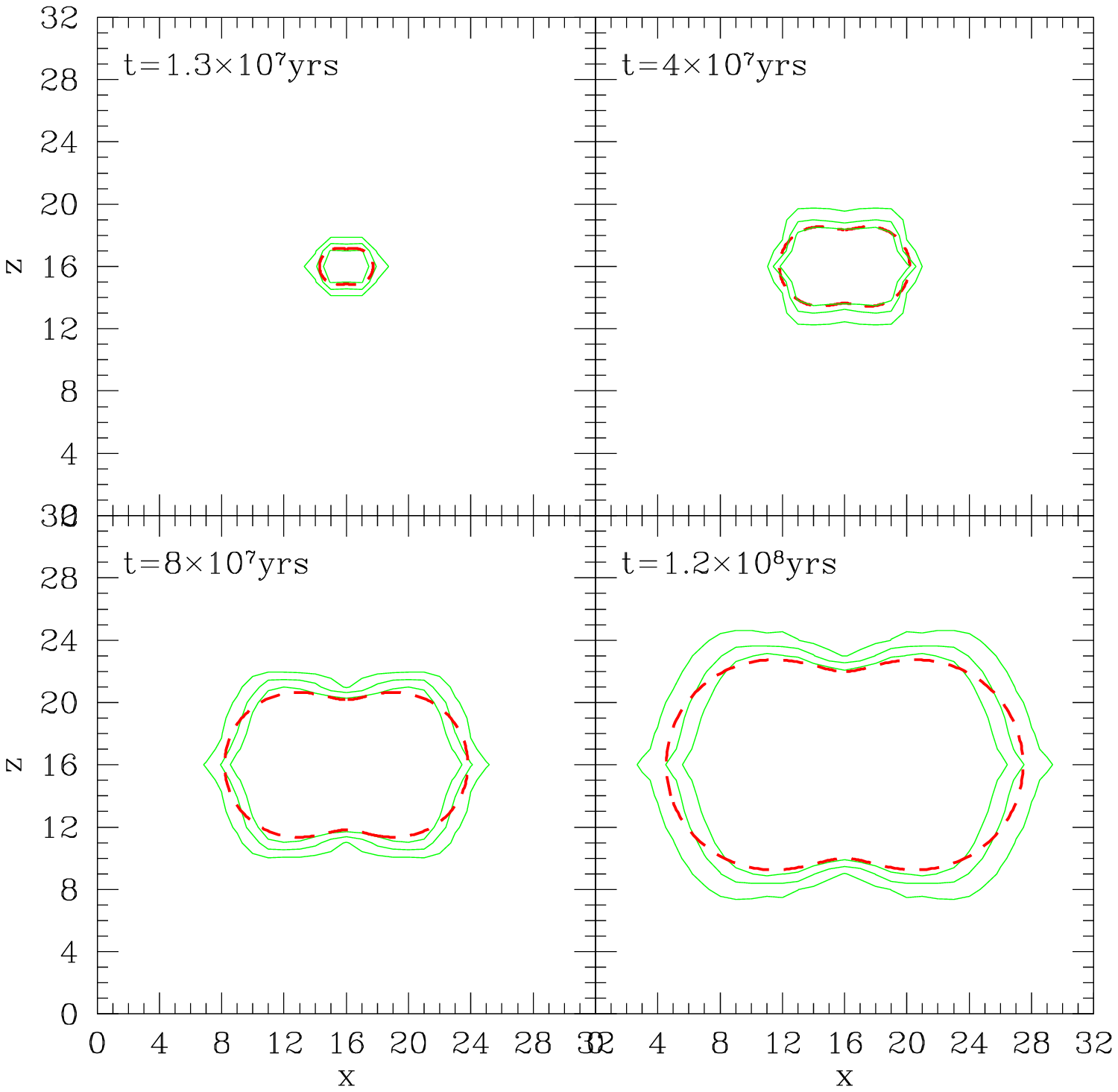}
\caption{
Distribution of neutral hydrogen fraction in the x-z plane (with $y=16$)
for a source sitting at cell (16,16,16)
in an elliptical isothermal  density distribution 
of $n(r)\propto r^{-2} f(\theta)$ 
[see Equation 31 for the elliptical function of $f(\theta)$]
at four epochs,
$(1.3\times 10^7, 4\times 10^7, 8\times 10^7, 1.2\times 10^8)~$yrs.
The dashed contours are the analytic results
and solid contours are obtained with the present algorithm,
indicating the neutral hydrogen fractions of $0.3, 0.6, 0.9$ inside out.
}
\label{ELLIPN2}
\end{figure}

\begin{figure}
\plotone{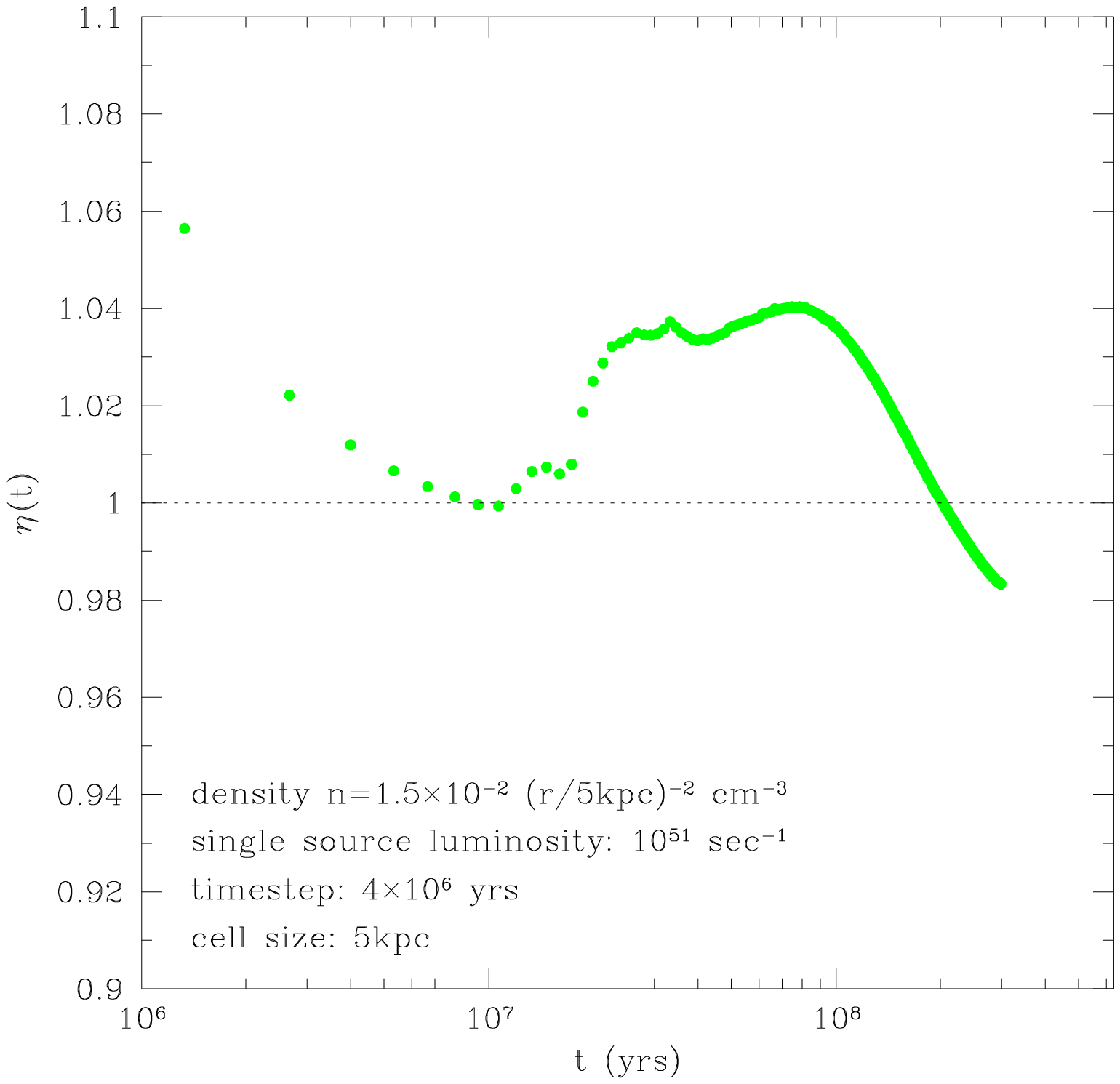}
\caption{
The error on photon number conservation as a function of time
for the case of an ionizing galaxy
surrounded 
an elliptical isothermal  density distribution
of $n(r)\propto r^{-2} f(\theta)$ 
(see Figure \ref{ELLIPN2}).
}
\label{ELLIPN2err}
\end{figure}

In realistic situations a source may 
sit inside a gas cloud that may not be spherical.
We examine an elliptical isothermal distribution
with the following density distribution 

\begin{equation}
n(r,\theta) = n_c (r/r_c)^{-2} {b\over\sqrt{1-(1-b^2)\cos^2\theta}},
\end{equation}

\noindent
where $\theta$ is the angle with respect to the
positive $z$ direction, 
and $b$ indicates the
ellipticity of the distribution.
The central cell at $r=0$ is designed not to be singular
but has a density equal to $n_c (\Delta x/r_c)^{-2}$,
where $\Delta x$ is the cell size.
The expected, orientation-dependent evolution of 
the radius of the ionization front follows Equation (30).

Figure \ref{ELLIPN2} 
shows the contours of the neutral hydrogen fraction in the x-z plane with $y=16$
at four epochs.
We use $\dot N_{ph}=10^{51}~$photon/sec, 
$n_c=1.5\times 10^{-2}~$cm$^{-3}$,
$r_c=5~$kpc 
and cell size of $\Delta x=5~$kpc.
Timestep used is $1.3\times 10^6~$yrs 
but the results are insensitive to the timestep.
It is seen that 
the algorithm handles the anisotropic density
distribution quite well and 
the agreement between the computed results
and analytic expectations is very good at all times,
with the computed ionization ``fronts"
tracing out the analytic results
quite nicely with an error no larger than about one cell.
The thickening of the computed ``fronts"
is real due to penetration of photons.
Figure \ref{ELLIPN2err} shows the degree
of conservation of photons.
It is seen that the number of photons is conserved at better than
$6\%$ with the average at about $1-4\%$.

\subsection{Ionization of a Sharply Divided Anisotropic Neutral Medium}

\begin{figure}
\plotone{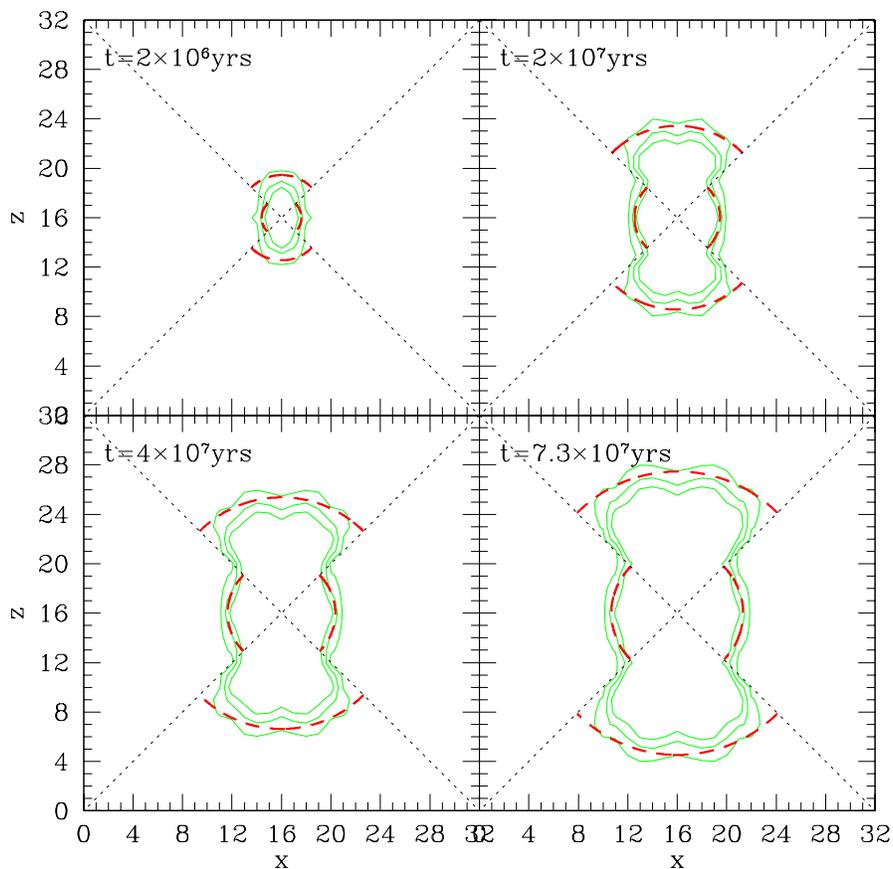}
\caption{
Distribution of neutral hydrogen fraction in the x-z plane (with $y=16$)
for a source sitting at cell (16,16,16) embedded
in a sharply divided anisotropic density field (see Equation 32)
at four epochs
$(2\times 10^6, 2\times 10^7, 4\times 10^7, 7.3\times 10^7)~$yrs.
The dashed contours are the analytic results
and solid contours are obtained with the present algorithm,
indicating the neutral hydrogen fractions of $0.3, 0.6, 0.9$ inside out.
}
\label{ANI}
\end{figure}

\begin{figure}
\plotone{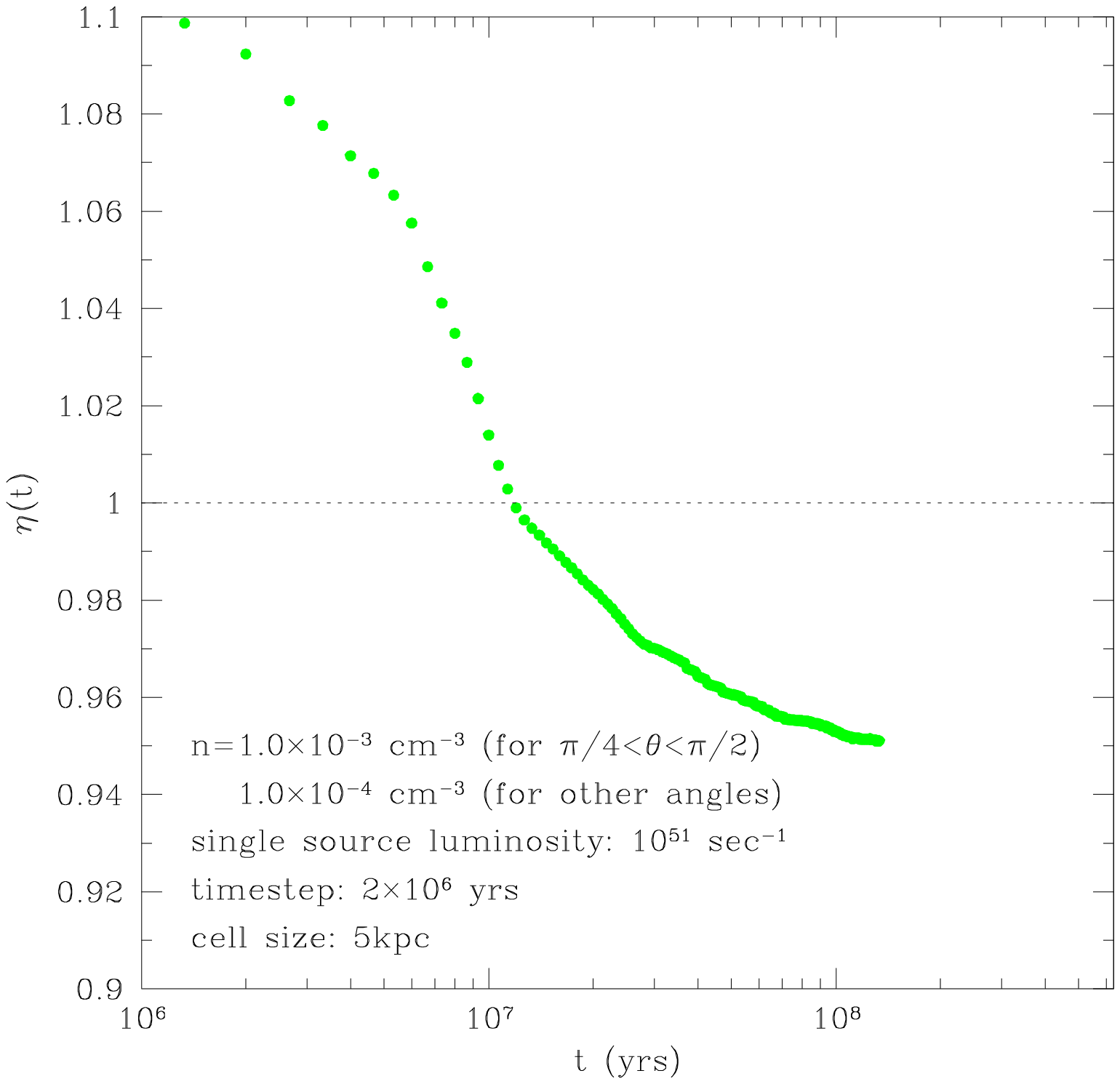}
\caption{
The error on photon number conservation as a function of time
for the case of an ionizing galaxy
surrounded by a sharply divided anisotropic density distribution
(see Figure \ref{ANI})
}
\label{ANIerr}
\end{figure}

So far we have examined several different cases with smooth 
density distributions.
It is prudent to test the algorithm 
in a more challenging situation.
We pick a case where there are
two different density fields separated by 
a sharp boundary.
An axisymmetric density field is prescribed as 
\begin{eqnarray}
n(\theta) &=& 10^{-3} {\rm cm^{-3}} {\ \ \ \rm for\ {\pi\over 4} \le \theta \le {3\pi\over 4}} \hfill \nonumber\\
&=& 10^{-4} {\rm cm^{-3}} {\ \ \ \rm otherwise.}\hfill
\end{eqnarray}

\noindent
The source sits at cell $(16,16,16)$.
The expected evolution of 
the radius of the ionization front follows Equation (28)
for each of the two domains with different 
speeds of the ionization fronts.

Figure \ref{ANI} 
shows the contours of the neutral hydrogen fraction 
in the x-z plane with $y=16$
at four epochs.
We use $\dot N_{ph}=10^{51}~$photon/sec, 
$r_c=5~$kpc and cell size of $\Delta x=5~$kpc.
Timestep used 
is $6.7\times 10^5~$yrs but the results are insensitive 
to the timestep.
The agreement between the computed results and the analytic results
are surprisingly good with the radius of the ionization front
accurate to about one cell.
At the boundaries separating the two regions
there is a slight over-shielding on the low
density side affected by the high optical depth, high density side.
The width of this ``affected" region is limited by
the size of the solid angle of each discretized angular element,
which in the present case is about $12^2$ square degrees.
Figure \ref{ANIerr} 
shows the degree
of conservation of photons,
indicating that the number of photons is conserved at better than
$10\%$ with the average at about $1-5\%$.

\subsection{Double Sources in a Uniform Density}

\begin{figure}
\plotone{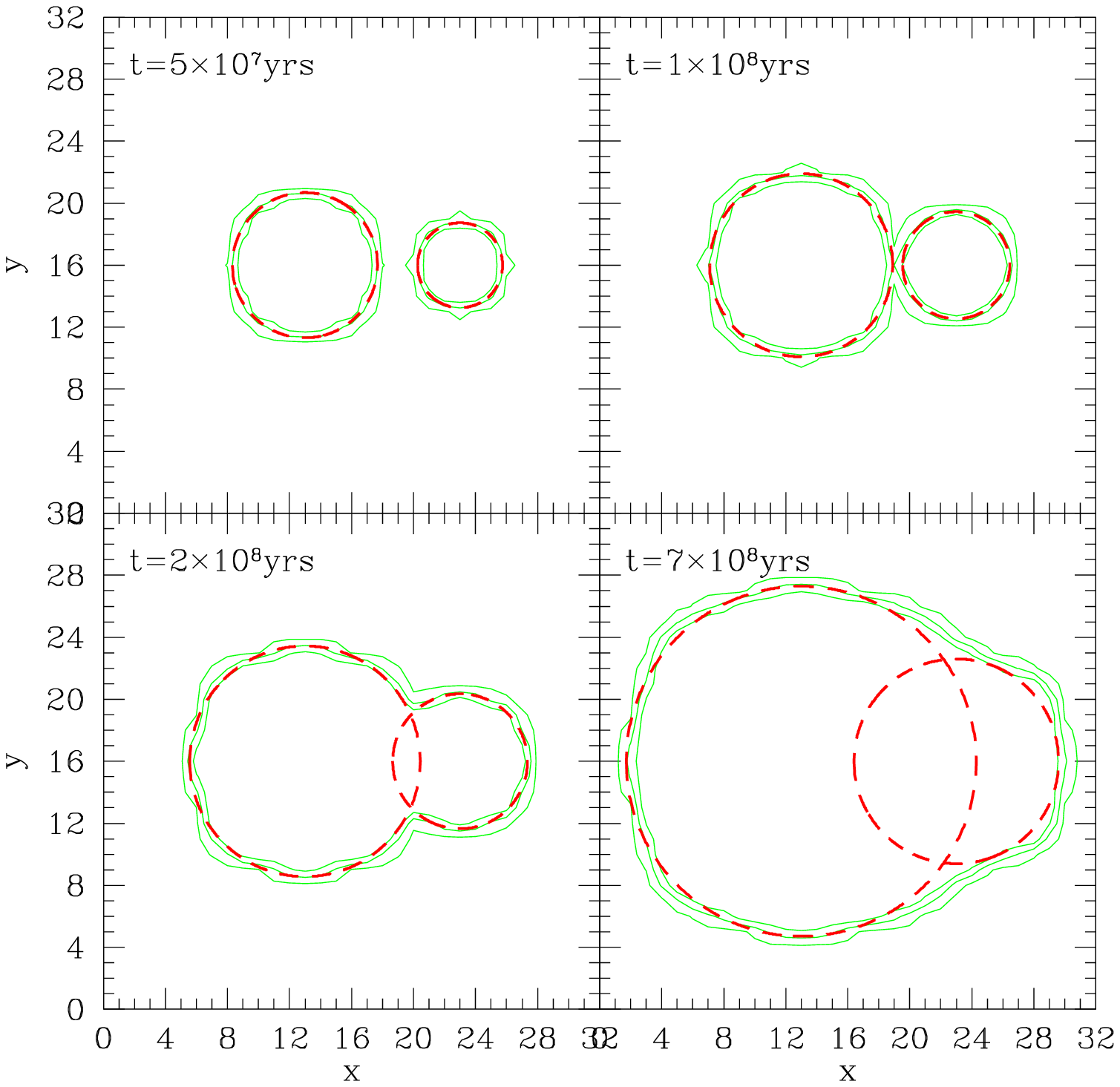}
\caption{
Distribution of ionization fraction in the x-y plane (with $z=16$)
for two static sources sitting at cell (23,16,16) 
and cell (13,16,16) with luminosities of 
$\dot N_{ph}=2\times 10^{50}~$photon/sec
and $\dot N_{ph}=10^{51}~$photon/sec, respectively,
with a separation of $50~$kpc,
at four epochs
$(5\times 10^7, 1\times 10^8, 2\times 10^8, 7\times 10^8)~$yrs.
The dashed contours are the analytic results
and solid contours are obtained with the present algorithm,
indicating the neutral hydrogen fractions of $0.3, 0.6, 0.9$ inside out.
}
\label{2GAL}
\end{figure}

\begin{figure}
\plotone{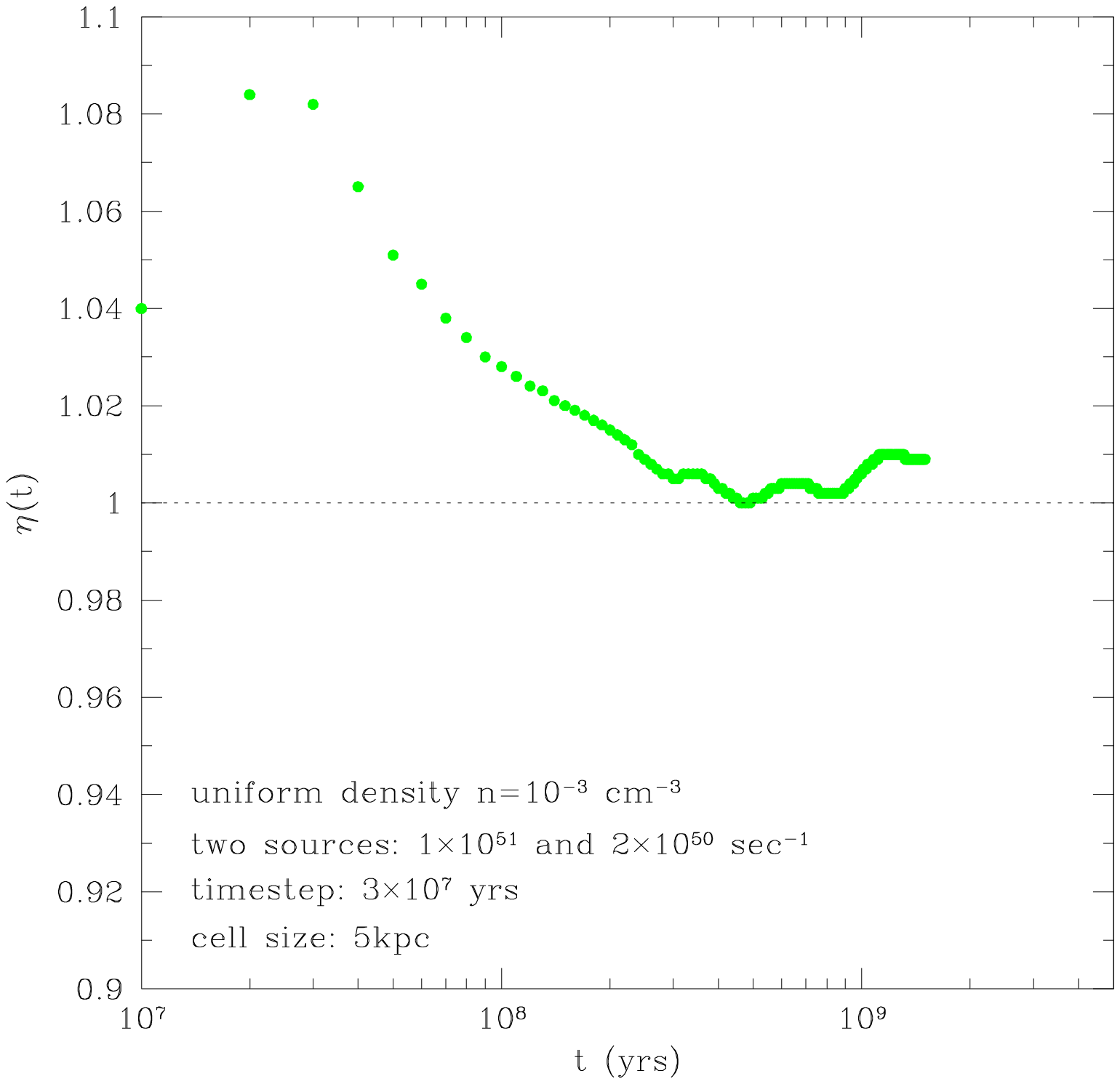}
\caption{
The error on photon conservation as a function of time
for the case of two ionizing galaxies with luminosities
of $10^{51}~$sec$^{-1}$ and $2\times 10^{50}~$sec$^{-1}$, respectively,
separated by $50~$kpc 
surrounded by a uniform density distribution with 
$n=1\times 10^{-3}~$cm$^{-3}$.
}
\label{2GALerr}
\end{figure}

Up to now we have tested cases with a single radiation source.
It is important that the algorithm is capable of 
handling interactions of multiple radiation sources
through the process of percolation, a process that is thought
to occur during the cosmological reionization.
We will demonstrate this by studying the situation
where there are two sources of unequal luminosities.
The two sources with luminosities $\dot N_{ph}=10^{51}~$photon/sec, 
and $2\times 10^{50}~$photon/sec, respectively,
sit in a uniform neutral medium of density $n=10^{-3}~$cm$^{-3}$ initially.
A cell size of $5~$kpc is used for the simulation box.
The two sources are located at cell (23,16,16)
and cell (13,16,16) with a separation of 10 cells (i.e., $50~$kpc).
Timestep used is $10^7~$yrs but the results are insensitive 
to the size of the timestep.

Prior to the overlap of the
two HII regions produced by the two sources separately,
the evolution of each HII region
is separate and the evolution of the radii of the ionization fronts
follows Equation (28).
Subsequent to the overlap their combined HII region 
will continue to expand but an analytic solution is not readily
available.
Figure \ref{2GAL} 
shows the contours of the neutral hydrogen fraction in the x-y plane with 
$z=16$
at four epochs.
It is clear that the algorithm nicely treats
the two separate regions without any interference before the overlap,
with the two separate ionization fronts traveling at the correct
speeds.
For the post-overlap era we only show the analytic  
solution as if overlap has not occurred for the sole purpose
of guiding the reader's eye.
The evolution of the ionization front in the post-overlap
era is more complicated but the computed results
appear to be quite reasonable,
with the overlapped region continuing
to expand and becoming rounder with time.
In this case, one is more keen to check if the total number of 
photons is conserved.
Figure \ref{2GALerr} 
shows the degree
of conservation of photons.
It is seen the number of photons is conserved at better than
$9\%$ with the average at about $1-2\%$.

\subsection{Quadruple Sources in a Uniform Density}

Let us make the situation a bit more interesting.
We will add two more sources to the case tested in \S 3.8.
The two additional sources have  
luminosities $\dot N_{ph}=3\times 10^{50}~$photon/sec, 
and 
$\dot N_{ph}=5\times 10^{50}~$photon/sec, 
sitting at cells (20,8,16) and (8,24,16)
but turned on with lags of $1\times 10^8~$yrs and 
$2\times 10^8~$yrs (relative to the turn-on time of the initial two sources),
respectively.

\begin{figure}
\plotone{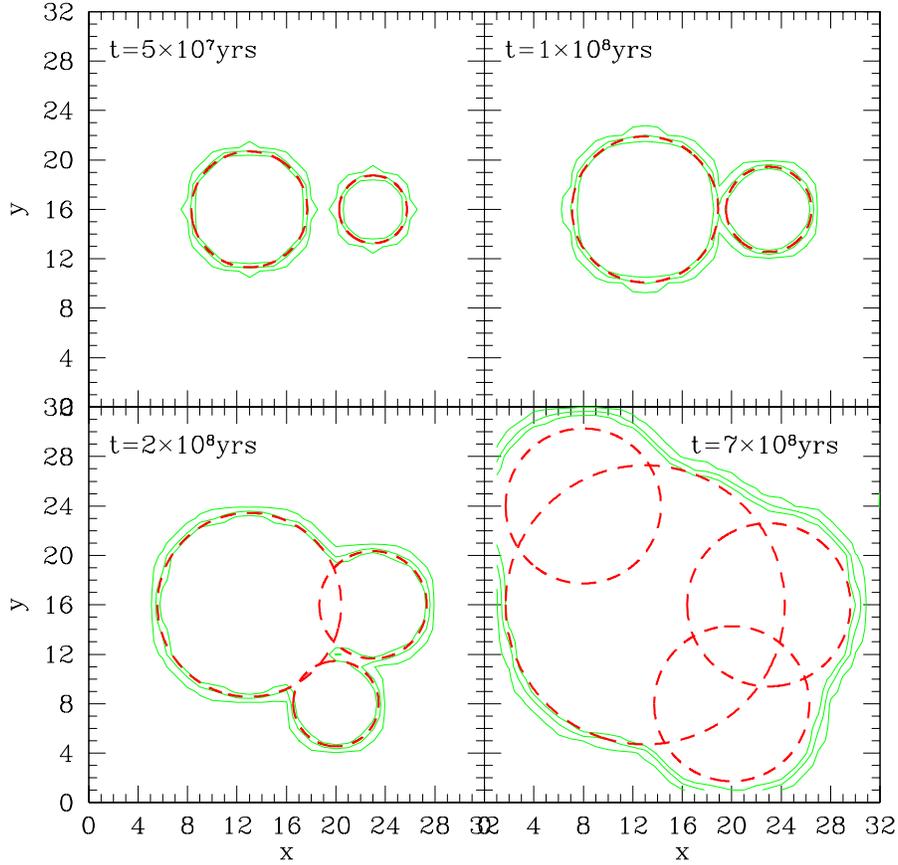}
\caption{
Distribution of neutral hydrogen fraction in the x-y plane (with $z=16$)
at four epochs
$(5\times 10^7, 1\times 10^8, 2\times 10^8, 7\times 10^8)~$yrs.
for four sources sitting at cells (23,16,16), (13,16,16),
(20,8,16) and (8,24,16),
with luminosities of 
$\dot N_{ph}=(2\times 10^{50}, 10^{51}, 3\times 10^{50}, 5\times 10^{50})~$
photon/sec, 
with turn-on times
at 
$t=(0, 0, 1\times 10^{8}, 2\times 10^{8})~$yrs,
respectively.
The dashed contours are the analytic results (only valid 
times before the HII regions overlap)
and solid contours are obtained with the present algorithm,
indicating the neutral hydrogen fractions of $0.3, 0.6, 0.9$ inside out.
}
\label{4GAL}
\end{figure}

\begin{figure}
\plotone{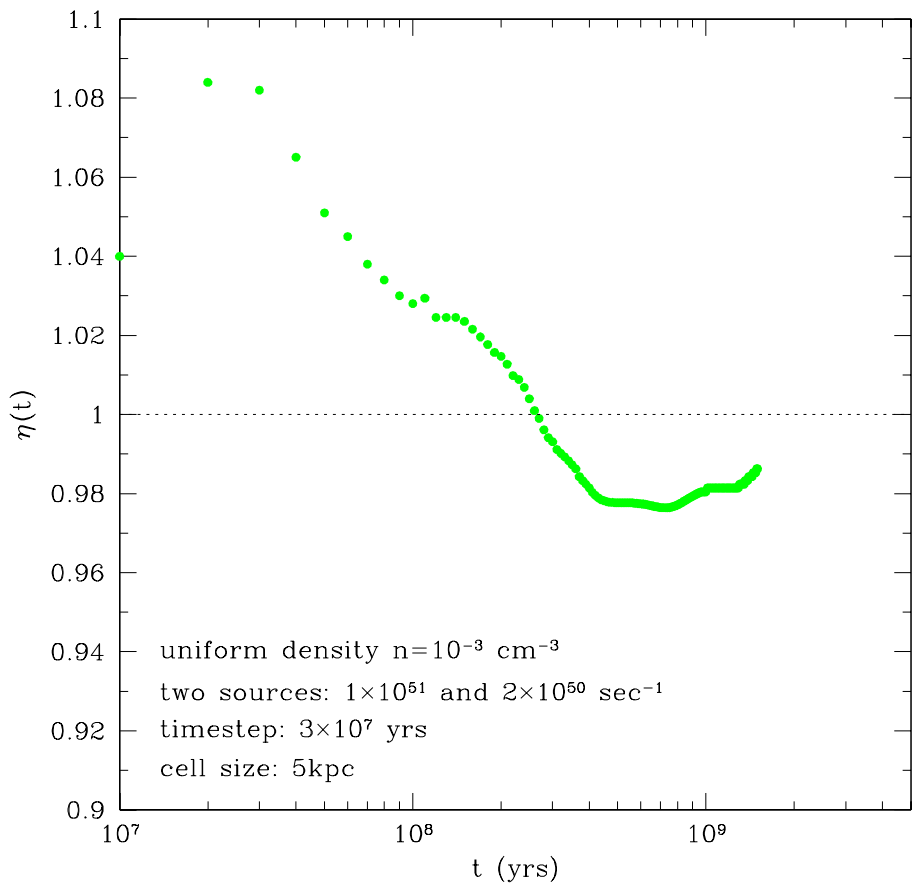}
\caption{
The error on photon number conservation as a function of time
for the case of four ionizing galaxies 
with luminosities of 
$\dot N_{ph}=(2\times 10^{50}, 10^{51}, 3\times 10^{50}, 5\times 10^{50})~$
photon/sec, 
with turn-on times at 
$t=(0, 0, 1\times 10^{8}, 2\times 10^{8})~$yrs,
respectively,
surrounded by a uniform density distribution with 
$n=1\times 10^{-3}~$cm$^{-3}$ (see Figure \ref{4GAL}).
}
\label{4GALerr}
\end{figure}

Figure \ref{4GAL} 
shows the contours of the neutral hydrogen fraction in the x-y plane with 
$z=16$ at four epochs.
For the post-overlap era we only show the analytic  
solution as if overlap has not occurred for the sake of illustration.
We see that the algorithm follows the separate HII regions, as expected.
Continuous formation of galaxies in time can evidently
be properly followed.
Subsequent mergers of HII regions occur as expected.
Figure \ref{4GALerr} 
shows the degree
of conservation of photons.
It is seen the number of photons is conserved at better than
$9\%$ with the average at about $1-3\%$.
The evolution of the ionization front in the post-overlap
era is very complicated but the computed results
look reasonable,
with the overlapped region continuing 
to expand in a fashion that is expected.
The fact that photon number is well conserved
and flux is designed to travel in the right direction
ensures that the computed results should be 
accurate.

\subsection{Shadowing by a Dense Gas Cloud}

\begin{figure}
\plotone{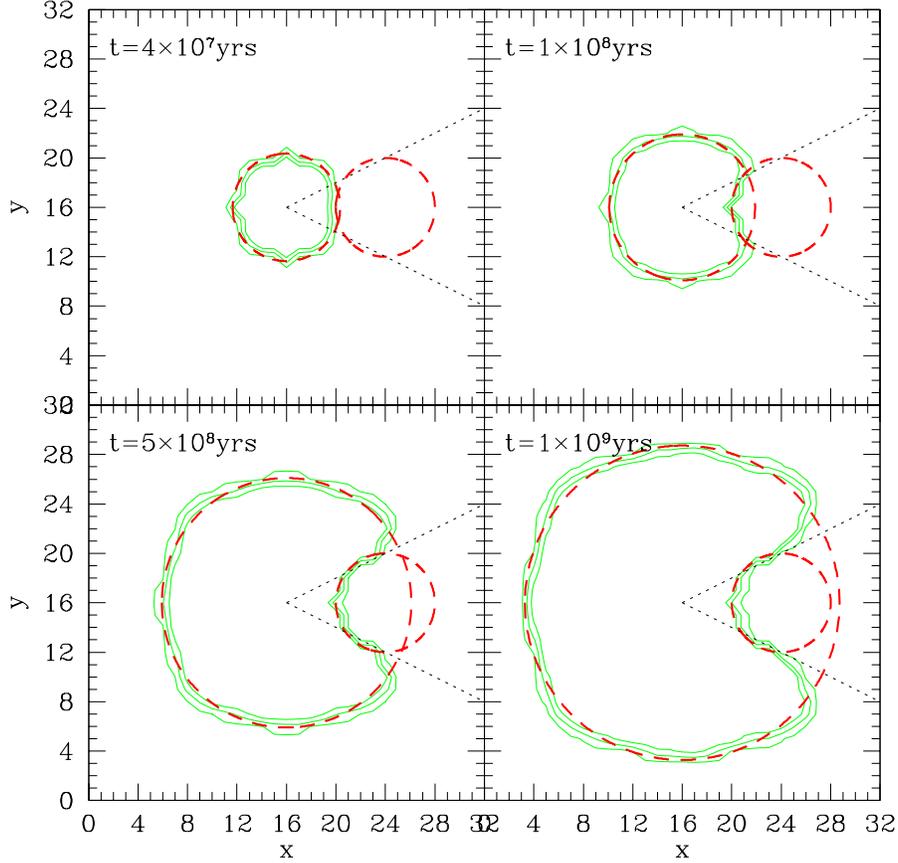}
\caption{
Distribution of neutral hydrogen fraction in the x-y plane (with $z=16$)
for a source sitting at cell (16,16,16)
of luminosity $10^{51}~$sec$^{-1}$
surrounded by a uniform density distribution with 
$n=1\times 10^{-3}~$cm$^{-3}$
at four epochs
$(4\times 10^7, 1\times 10^8, 5\times 10^8, 1\times 10^9)~$yrs.
In addition, there is a spherical gas cloud
of radius $20~$kpc centered at $(24,16,16)$ 
with a density distribution of 
$0.1(r/20 {\rm kpc})^{-1}$.
The small dashed circle on the right in each panel
indicates the size of the halo.
The (larger) dashed contours (on the left) are the analytic results
and solid contours are obtained with the present algorithm,
indicating the neutral hydrogen fractions of $0.3, 0.6, 0.9$ inside out.
The analytic results are only valid before the ionization
front reaches the halo; 
at subsequent times regions behind the halo 
delimited by the two dotted lines 
should remain neutral 
and the (large) circle on the left serves only to indicate
the position of the ionization front if there were no obscuration.
}
\label{HALO}
\end{figure}

\begin{figure}
\plotone{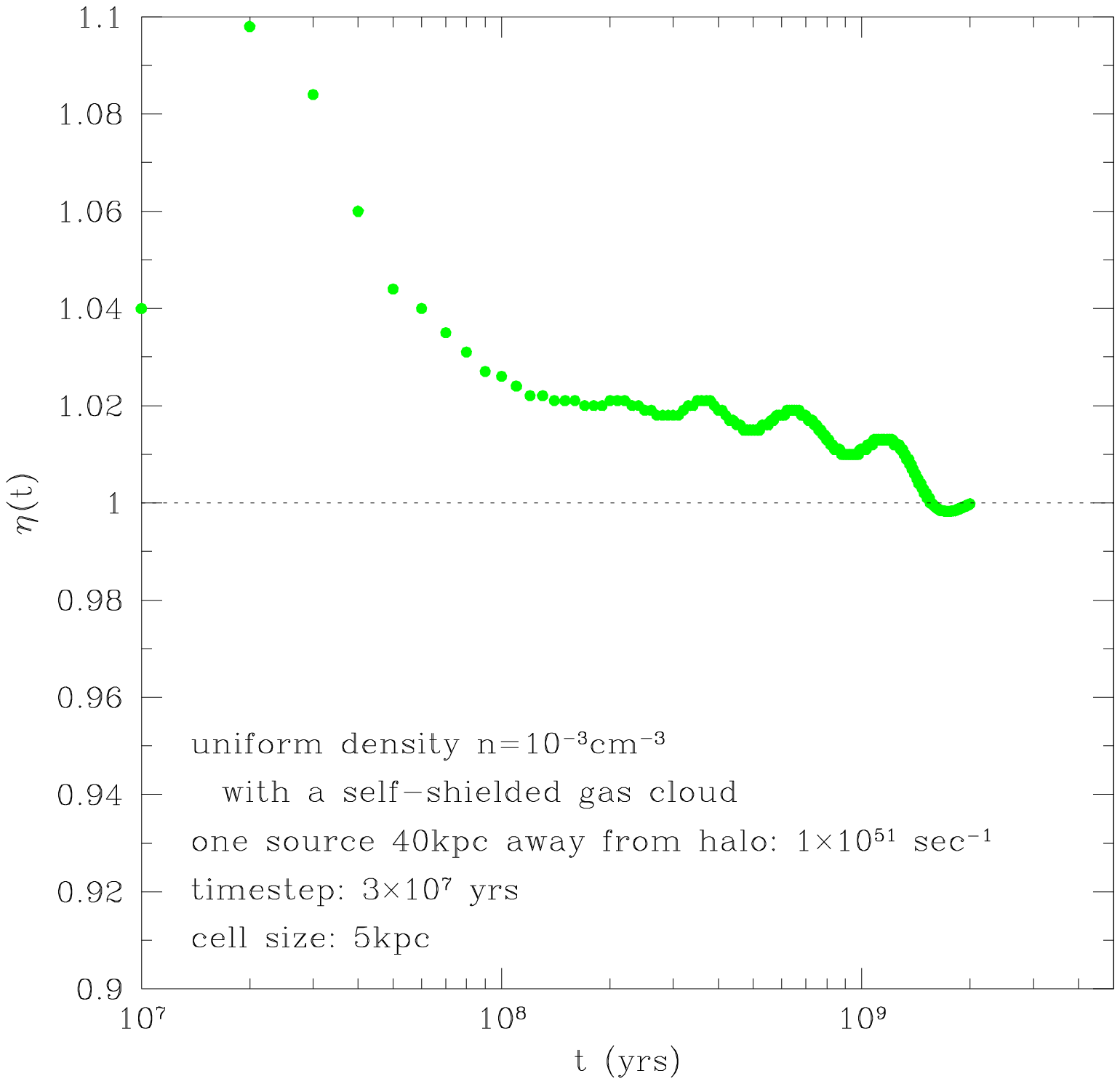}
\caption{
The error on photon number conservation as a function of time
for the case of one ionizing galaxy with luminosity
of $10^{51}~$sec$^{-1}$
surrounded by a uniform density distribution with 
$n=1\times 10^{-3}~$cm$^{-3}$.
In addition, there is a spherical gas cloud
of radius $20~$kpc centered at $(24,16,16)$ 
with a density distribution of 
$1\times 10^{-1} (r/20 {\rm kpc})^{-1}$.
}
\label{HALOerr}
\end{figure}

We will now step back to a case with a single source.
But we will include in the uniform density field
a dense gas cloud, which is designed to be optically
thick to cast a shadow in the direction of radiation propagation.
It is expected that shadowing should be common in the real
universe since galaxies and other dense clouds (such as damped
Lyman alpha systems)
are (at least partially) opaque to radiation.
A spherical gas cloud 
of radius $20~$kpc with a density distribution of 
$1\times 10^{-1} (r/20{\rm kpc})^{-1}~$cm$^{-3}$ is centered 
at cell $(24,16,16)$.
The rest of the simulation box has a uniform density
of $1\times 10^{-3}~$cm$^{-3}$.
The ionizing source is located at cell (16,16,16)
with a luminosity of $10^{51}~$sec$^{-1}$.
A cell size of $\Delta x=5~$kpc is used for the simulation.
Figure \ref{HALO} 
shows the neutral hydrogen fraction contours
at four epochs.
The analytic results are only valid before the ionization
front reaches the halo; 
at subsequent times regions behind the halo 
within the region delimited by the two dotted lines 
should remain neutral. 
We indeed see the shadow
cast by the dense gas cloud.
The slightly overshadowing near the edges 
of the shadowed region is due to the limited
resolution of angular discretization, also seen
in Figure \ref{ANI}. 

The relatively complicated situation here does not 
guarantee that photon number conservation will be obeyed.
Figure \ref{HALOerr} 
shows the degree
of conservation of photons.
Quite reassuringly, we see 
that the number of photons is conserved at better than
$10\%$ with the average at about $1-2\%$.

\subsection{Outside-In Ionization of a Spherical Isothermal Gas Cloud}

Let us next examine another class of ionization processes,
where external diffuse radiation ionizes isolated
overdense regions.
This type of ionization process 
is common in cosmological situations where 
low density regions become ionized first.
In this case we check the photon number conservation
in a slightly different way as:

\begin{equation}
\eta(t) = {N_e(t)\over N_{incoming}(t)},
\end{equation}

\noindent
where $N_e(t)$ is the number of free electrons created by time $t$
in the simulation box
and $N_{incoming}$ is the number of incoming photons 
from the diffuse background that are absorbed 
in the simulation box volume by time $t$.
If photon number conservation is strictly observed,
$\eta$ would be unity.
Recombination time is set to infinity.

The first simple example is the ionization 
of a spherical isothermal gas cloud of 
density $n = n_c (r/r_c)^{-2}~$cm$^{-3}$
with an outer cutoff radius $r_i$;
the central cell at $r=0$ has a density 
equal to $n_c (\Delta x/r_c)^{-2}$ to avoid singularity,
where $\Delta x$ is the cell size.
The expected evolution of 
the radius of the ionization front, which now travels
inward towards the center of the gas cloud, is approximately:

\begin{equation}
r(t) = {r_i n_c r_c^2 \over 0.5F r_i t + n_c r_c^2},
\end{equation}

\noindent
where $F$ is the diffuse flux assumed to be constant with time.
Note that the factor $0.5$ in front of flux $F$ in Equation (34)
is due to the fact that any element on the surface 
is subject to radiation only at half the total solid angle.

\begin{figure}
\plotone{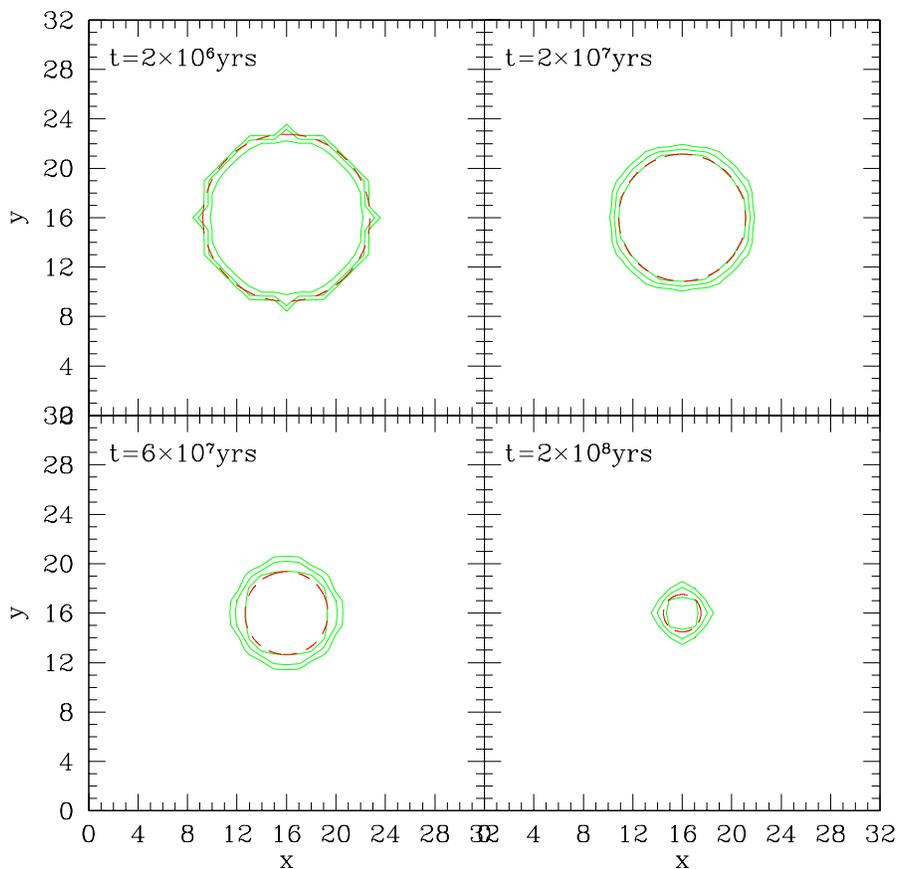}
\caption{
Distribution of neutral hydrogen fraction in the x-y plane (with $z=16$)
for an isothermal gas cloud sitting at cell (16,16,16),
exposed to an external isotropic diffuse radiation background,
at four epochs
$(2\times 10^6, 2\times 10^7, 6\times 10^7, 2\times 10^8)~$yrs.
The dashed contours are the analytic results (Equation 34)
and solid contours are obtained with the present algorithm.
indicating the neutral hydrogen fractions of $0.3, 0.6, 0.9$
outside in.
}
\label{ISLAND}
\end{figure}

\begin{figure}
\plotone{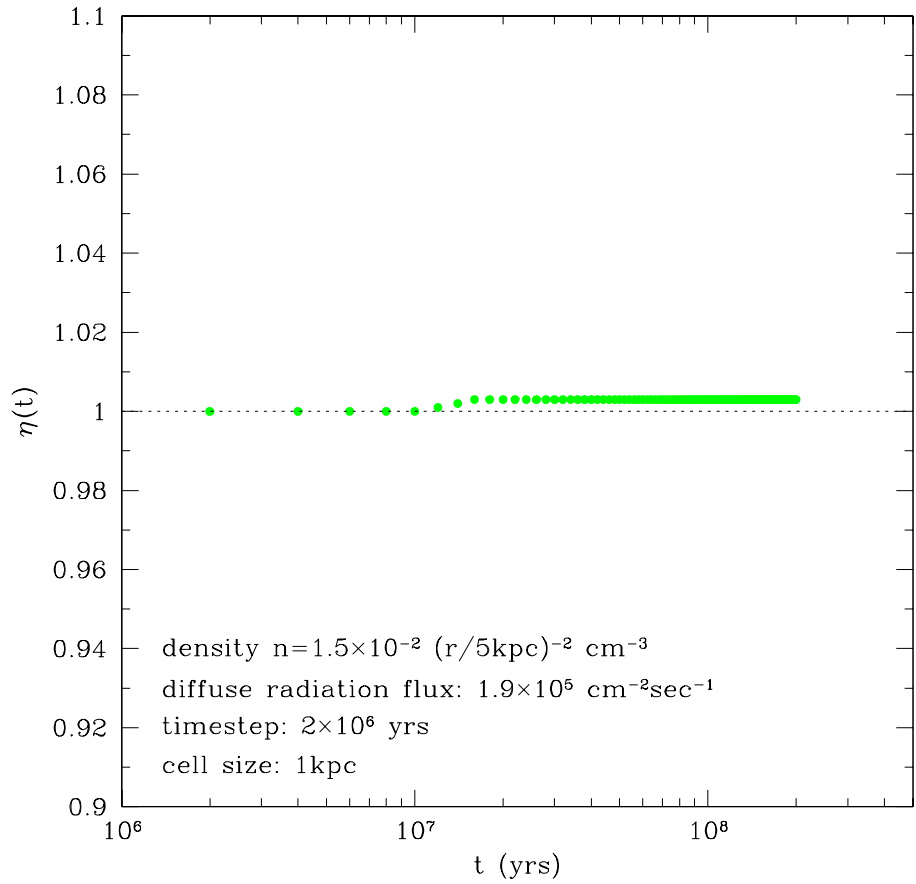}
\caption{
The error on photon number conservation as a function of time
for the case of a halo 
of density $n_c=1.5\times 10^{-2}(r/5{\rm kpc})^{-2}~$cm$^{-3}$,
subject to a diffuse background flux
of $19\times 10^{5}~$cm$^{-2}$sec$^{-1}$.
}
\label{ISLANDerr}
\end{figure}

\begin{figure}
\plotone{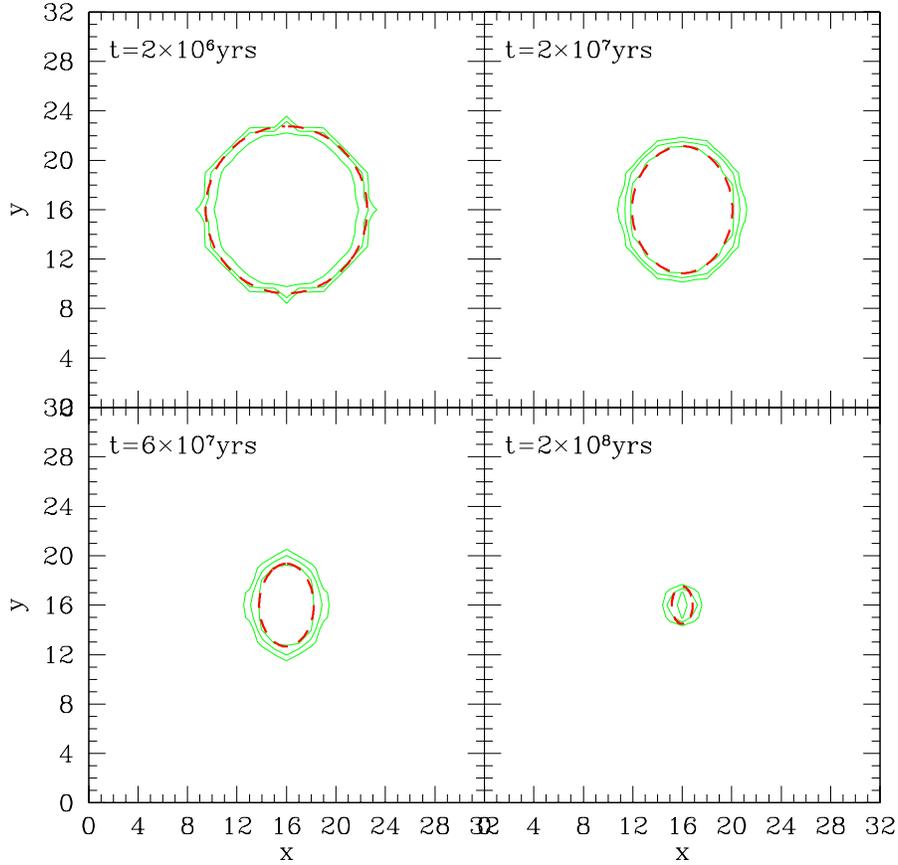}
\caption{
Distribution of neutral hydrogen fraction in the x-z plane with $y=16$ 
for an elliptical isothermal gas cloud (see Equation 35)
sitting at cell (16,16,16),
exposed to an external isotropic diffuse radiation background,
at four epochs
$(2\times 10^6, 2\times 10^7, 6\times 10^7, 2\times 10^8)~$yrs.
The dashed contours are the analytic results
and solid contours are obtained with the present algorithm,
indicating the neutral hydrogen fractions of $0.3, 0.6, 0.9$
outside in.
}
\label{ELLIPISLAND}
\end{figure}

\begin{figure}
\plotone{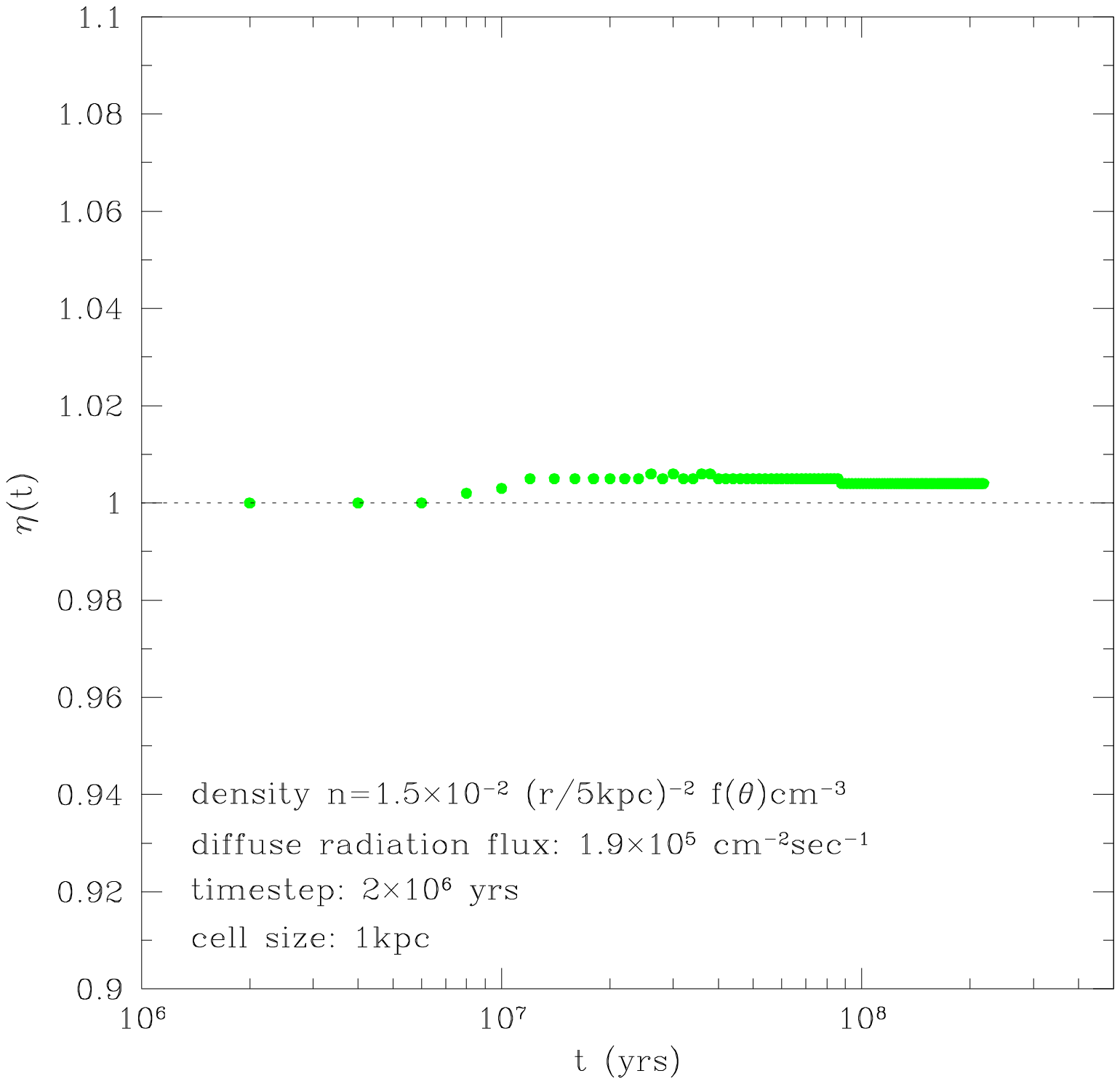}
\caption{
The error on photon number conservation as a function of time
for the case of an elliptical isothermal gas cloud
(see Equation 35)
sitting at cell (16,16,16),
exposed to an external isotropic diffuse radiation background.
}
\label{ELLIPISLANDerr}
\end{figure}

Figure \ref{ISLAND} shows the evolution of the 
ionization front in the x-y plane with $z=16$.
We use $r_c=5~$kpc, $n_c=1.5\times 10^{-2}$, $\Delta x=1~$kpc,
$b=0.5$ and $F=1.9\times 10^5~$cm$^{-2}$sec$^{-1}$.
We see that the analytic evolution of the ionization front
is well tracked by the computed results with 
error on the radius no larger than one cell.
The spiky surface at the earliest time shown
reflects the initial condition laid out on a Cartesian grid.
Figure \ref{ISLANDerr} shows the photon number conservation
as a function of time and again indicates
that the method conserves total number of photons very well
at $<1\%$.
The high degree of photon conservation in case is attributable
to the fact that there is no obscuration 
for the cells on the ionization surface that receive ionizing
photons.

\subsection{Outside-In Ionization of an Elliptical Isothermal Gas Cloud}

We next consider ionization 
of an isolated elliptical isothermal gas cloud of the following density profile 
by a diffuse ionizing background 

\begin{equation}
n(r,\theta) = n_c (r/r_c)^{-2} {b\over \sqrt{1-(1-b^2)\cos^2\theta}};
\end{equation}

\noindent
the central cell at $r=0$ has a density 
equal to $n_c (\Delta x/r_c)^{-2}$ to avoid singularity,
where $\Delta x$ is the cell size.
The expected evolution of 
the radius of the ionization front follows
Equation (34) with an angle-dependent $n_c$ indicated by Equation (35).

Figure \ref{ELLIPISLAND} shows the evolution of the 
ionization front in the x-z plane with $y=16$.
We use $r_c=5~$kpc, $n_c=1.5\times 10^{-2}$, $\Delta x=1~$kpc
and $F=1.9\times 10^5~$cm$^{-2}$sec$^{-1}$.
We see that the analytic evolution of the ionization front
is well traced by the computed results with 
error on the radius no larger than one cell.
Figure \ref{ELLIPISLANDerr} shows the photon number conservation
as a function of time and again indicates
that the method conserves total number of photons very well
at $<1\%$.

\subsection{Radiation Propagation In a Realistic Cosmological Density Field}

Finally, we turn to the radiation propagation
in a realistic density field produced by cosmological simulations.
In this case, the situation is much more complicated
and no analytic solution is possible.
But many of the cases tested above
have significant bearings on this realistic case.
We take a subbox from one of our latest simulations (Cen \etal 2001)
at redshift $z=6$.
The subbox has a $32^3$ grid and a cell size of 
$4.65~$kpc proper.

\begin{figure}
\plotone{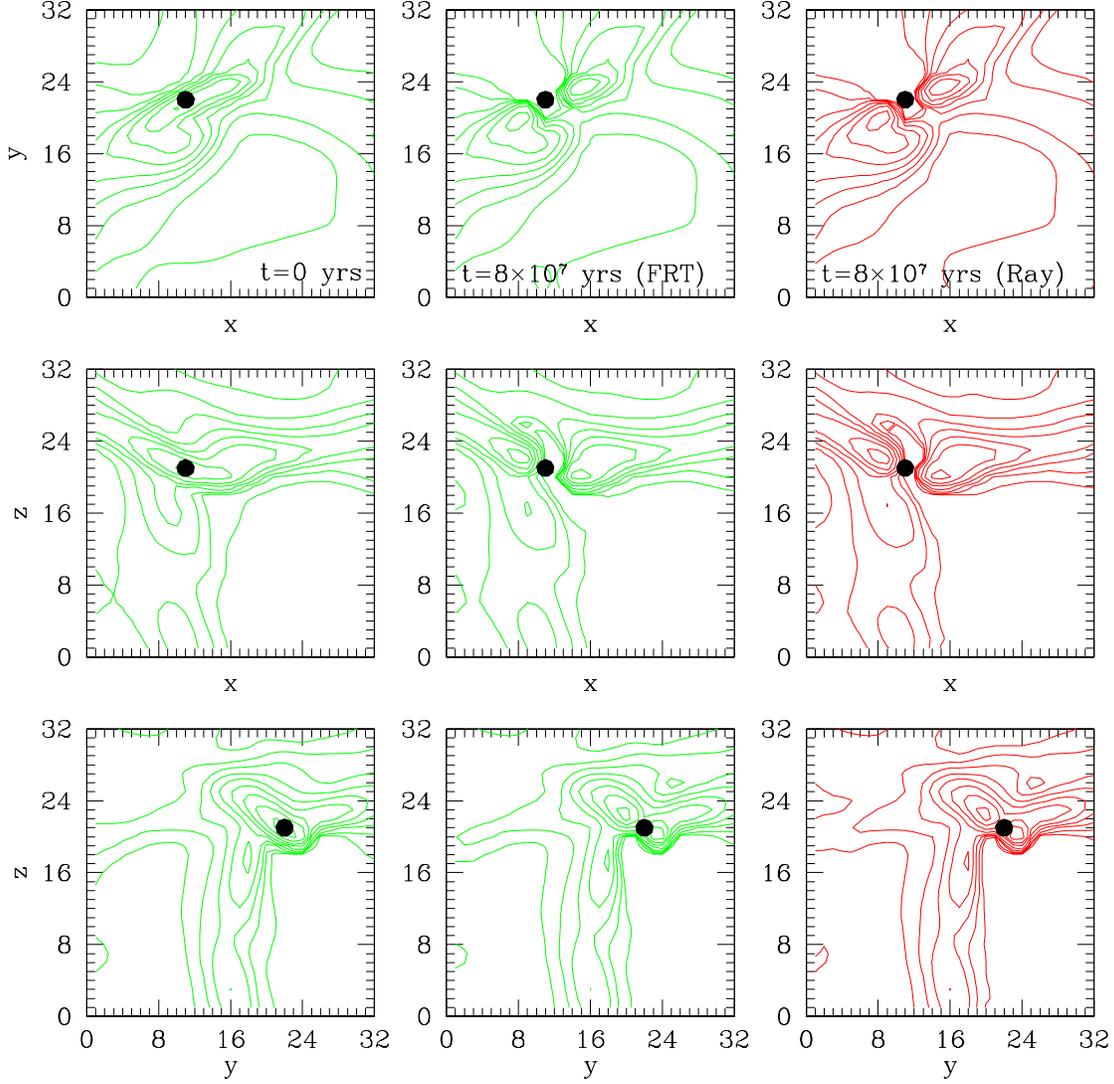}
\caption{
The left column shows projected gas density fields 
along three faces of the box (three rows)
for a three-dimensional box with $32^3$ grid points and a cell size of 
$4.65~$kpc proper at the initial time.
The solid dot indicates the position of the putative galaxy of
luminosity $10^{51}~$sec$^{-1}$.
The contours are: the first contour is at the mean density
and subsequent ones increase by 0.2 dex per contour.
The middle column shows 
projected neutral hydrogen density fields along three faces
of the box 
at $t=8\times 10^7$yrs, computed with the present method.
The right column shows 
projected neutral hydrogen density fields along three faces
of the box 
at $t=8\times 10^7$yrs, computed using a high
resolution ray tracing code (with 38416 rays from the source).
}
\label{REAL1}
\end{figure}

\begin{figure}
\plotone{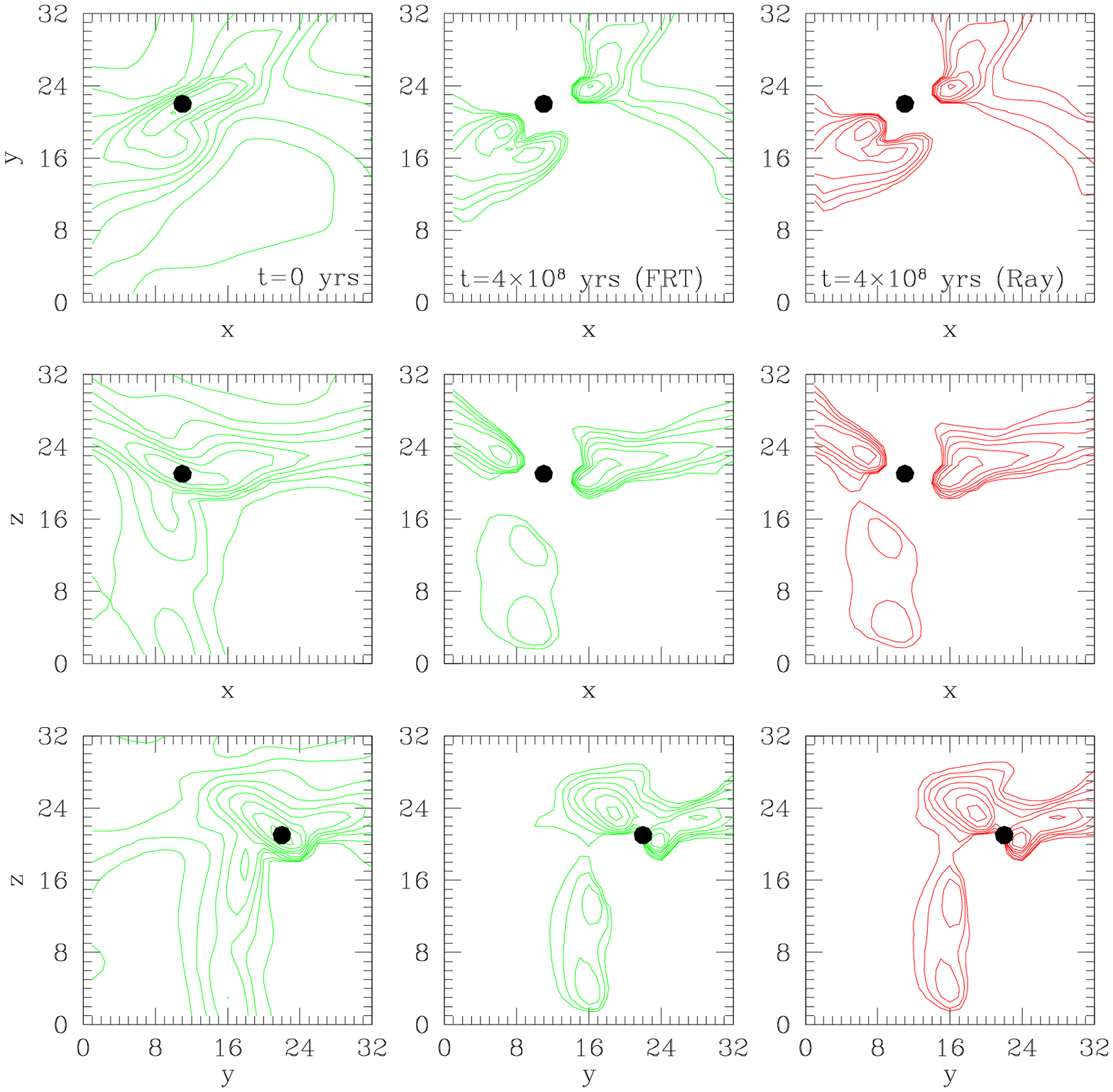}
\caption{
The left column shows projected gas density fields 
along three faces of the box (three rows)
for a three-dimensional box with $32^3$ grid points and a cell size of 
$4.65~$kpc proper at the initial time.
The solid dot indicates the position of the putative galaxy of
luminosity $10^{51}~$sec$^{-1}$.
The contours are: the first contour is at the mean density
and subsequent ones increase by 0.2 dex per contour.
The middle column shows 
projected neutral hydrogen density fields along three faces
of the box 
at $t=4\times 10^8$yrs, computed with the present method.
The right column shows 
projected neutral hydrogen density fields along three faces
of the box 
at $t=4\times 10^8$yrs, computed using a high
resolution ray tracing code (with 38416 rays from the source).
}
\label{REAL2}
\end{figure}

\begin{figure}
\plotone{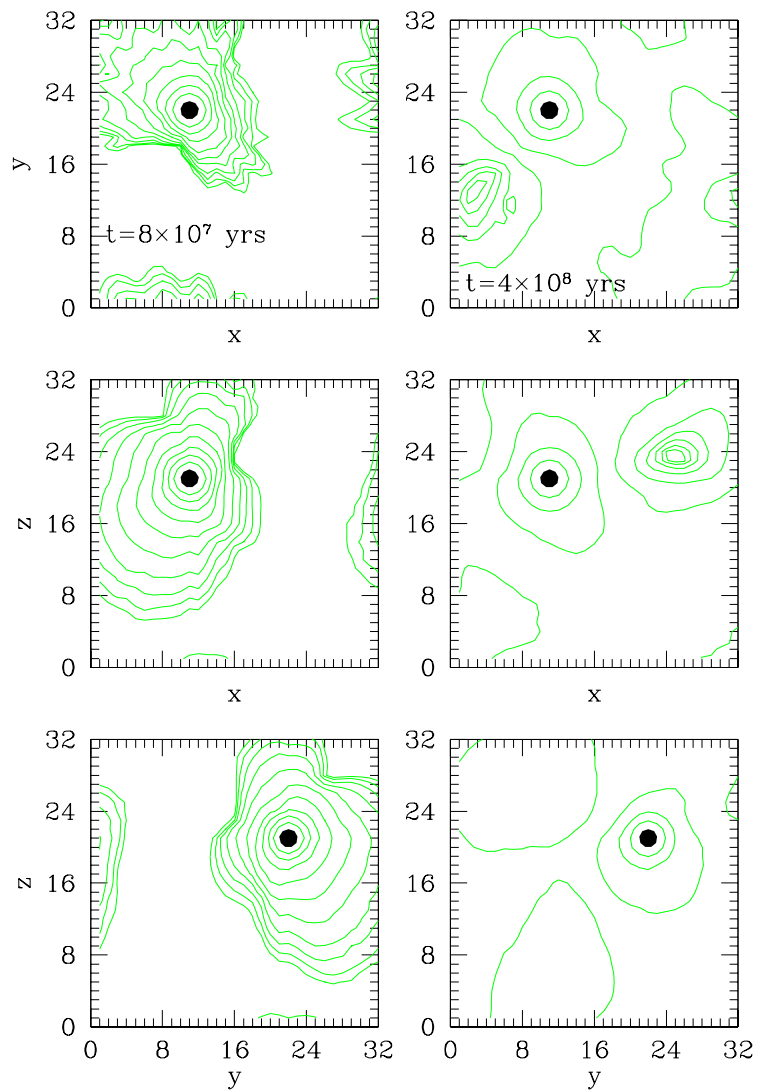}
\caption{
Volume-weighted flux distributions projected along three faces of the box 
at the two times,
$t=(8\times 10^7, 4\times 10^8)$yrs,
shown in 
Figure \ref{REAL1} 
and Figure \ref{REAL2}, respectively.
The first contour around the solid dot (the galaxy)
has a flux of $10^{8.5}$cm$^{-2}$sec$^{-1}$ and
the decrement for the successive contours is 0.25 dex.
}
\label{REALflux}
\end{figure}

\begin{figure}
\plotone{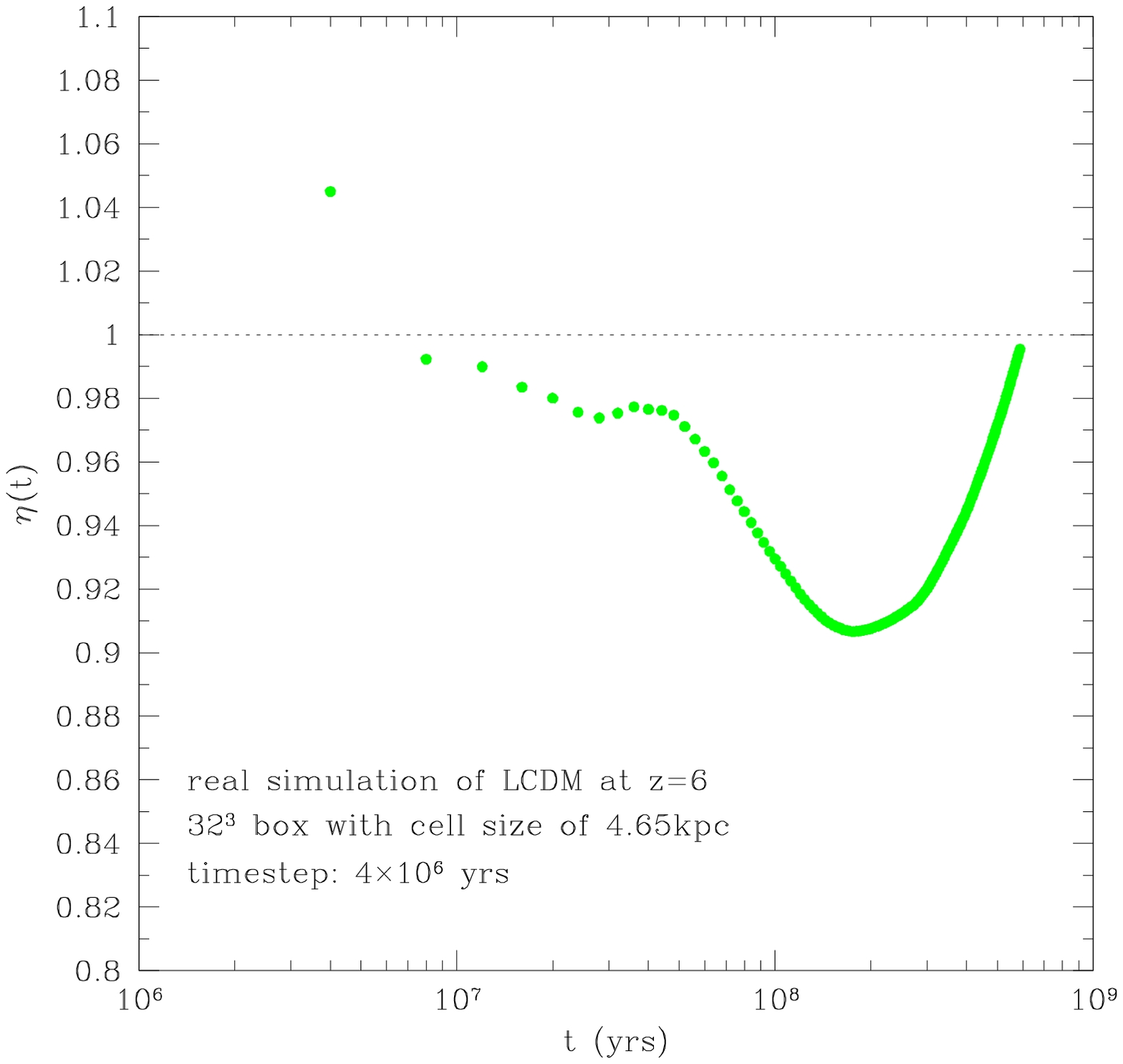}
\caption{
The error on photon number conservation as a function of time
for the case of ionization of
a realistic density field produced in cosmological simulations
at $z=6$.
}
\label{REALerr}
\end{figure}

We turn on by hand a galaxy of ionizing luminosity
of $\dot N_{ph}=10^{51}~$sec$^{-1}$
at cell $(11,22,21)$ corresponding to 
the densest cell in the box,
as indicated by the solid dot in Figures (24,25,26).
We assume that the gas is composed entirely of hydrogen
and at the starting time hydrogen is entirely neutral.
For simplicity we have set the recombination time to infinity.
It is noted that the galaxy sits 
in a generic filament, which 
is embedded in a complex structure containing voids and other
interesting structures.
Since the subbox is a small region of a large, periodic box,
it is not periodic itself.
But we treat it as if it were periodic.

Figure \ref{REAL1} 
and Figure \ref{REAL2} 
show the distribution of neutral hydrogen
at time $t=(8\times 10^7,4\times 10^8)$yrs (middle columns),
respectively, compared to that at the starting time ($t=0$; left column).
The most notable feature is that the radiation propagation
is highly anisotropic.
The ionizing photons quickly clear out 
widening tunnels towards low density (void) regions in the direction
perpendicular to the filament,
while radiation front travels at much slower speeds in other directions.
Although this is not at all surprising,
the figures  do remind us of the complex situations
encountered in cosmological structures and
re-iterate the need for a proper treatment of radiative transfer.
We also use a high resolution ray tracing code 
to obtain numerically the approximate ``true" distribution,
shown as the right columns.
Note that in the ray tracing code one just counts photons
and compares to the number of hydrogen atoms along each ray cone;
photon penetration ahead of the ionization front
is not taken into account and ionization front is taken to be
abruptly sharp.
Nevertheless, the results obtained with the ray tracing method
should be fairly close to the `truth".
We see that the agreement between the result obtained with
the present method and that with the ray tracing method
is excellent.

Figure \ref{REALflux} 
shows the projected flux distributions 
(volume-weighted) at the two epochs respectively.
We see that flux distributions are quite complex
but consistent with the neutral hydrogen
distributions. It is seen at the earlier epoch,
when substantial blocking or shadowing exists,
the contours of the flux distribution 
tend to orient perpendicular to the contours
of the neutral hydrogen density.
Note that in the present calculations we have set the recombination
time to infinity; in a more involved calculation including 
recombination it is expected that fluctuations in the 
flux distributions would be enhanced.

Finally, Figure \ref{REALerr} 
shows how the total number of photons is conserved.
We are once again seeing satisfactory conservation of photons
in this realistic simulation 
at about $1-10\%$ level with 
a time averaged accuracy of $3-5\%$.
This level of photon number conservation is quite adequate for
cosmological simulations,
since other uncertainties, including galaxy luminosity,
radiation escape fraction from the galaxies,
star formation efficiency, etc., 
are usually much larger.
Besides, we expect that we may be 
able to improve the accuracy of the algorithm 
by more detailed experimenting and careful fine tuning,
if necessary.

The fact that photon number is well conserved
and flux is designed 
under the algorithm 
to propagate in the right directions
guarantee that the results should be reliable,
as verified by the comparison with the results from
a high resolution ray tracing method.

\subsection{Summary on the Tests}

We have performed a variety of tests on the proposed 
algorithm for transferring radiation in three dimensional
space. 
The tests are designed to be relevant to 
cosmological applications.
It is worthwhile to re-emphasize that 
the proposed algorithm, by design,
guarantees that flux propagates in the right direction
in any situation.
Thus, the tests should mainly demonstrate how accurately
the amplitude of the flux is computed.

To summarize the results
we find,
the flux is computed very accurately,
resulting in ionization fronts that 
travel at the correct speed 
with error no larger than one cell in all cases.
Photon number is conserved
with a maximum error of about $10\%$
and an average error of $1-5\%$ over hundreds
of time steps for all the cases tested,
with $m_a=256$ angular elements and 
$m_r=18$ logarithmically spaced radial bins on a $32^3$ grid.
We note that in the limit of infinite values for $m_r$ and $m_a$,
the present method solves
the equation of the radiative transfer exactly
and the approximation would represent the truth.
In other words, the tests have basically shown
that with reasonable number of angular and radial discretization elements
the proposed method gives quite accurate results.

Finally, we find that the accuracy of the algorithm
does not depend sensitively on the time step used.
A time step of size comparable to the typical time step
in cosmological hydrodynamic simulations
appears sufficient.
This feature is desirable with regard to the overall
computational cost.

\section{Discussion and  Conclusions} 

In light of the recent observations of high redshift quasars 
(e.g., Fan \etal 2001),
indicating that the cosmological reionization may be
just ending at redshift $z\sim 6$ 
(Becker \etal 2001; Barkana 2001; Cen \& McDonald 2001),
a new and exciting frontier is forming.
Inhomogeneous reionization of the universe
is now within the observational reach and could potentially
provide a great tool to test both cosmological models
and astrophysics of galaxy formation (e.g., Barkana \& Loeb 2001).
It has thus become mandatory to perform cosmological
hydrodynamic simulations that tackle this problem with
an adequate treatment of radiative transfer.
Furthermore, even in the post-reionization era,
an initial inhomogeneous reionization process
coupled with the significant
clustering of radiation sources (galaxies or quasars)
as well as the fluctuating density field
likely produces fluctuations both in the thermodynamic properties
of the cosmic gas and in the ionizing radiation field.
Such fluctuations may substantially alter quantitatively (or even
qualitatively) the picture of the optically thin regions of 
the \lya forest on a variety of scales, which would
limit our ability of accurately extracting important cosmological
information from observations of the \lya forest.
Lastly (but not the least), since almost all neutral gas resides and 
star formation occurs {\it only} 
in dense, optically thick regions,
many questions pertaining
neutral gas and star formation in the universe can be answered 
with confidence only when radiative transfer is properly included.

We have developed a fast, accurate and robust algorithm for 
radiative transfer in three-dimensional space.
The core of the algorithm is based on 
a formulation that the summation of any quantity 
(such as emissivity or opacity)
over any volume can be written 
in the standard convolution form,
which can be computed efficiently 
using the Fast Fourier Transform techniques.
We will name this method ``FRT" (Fast Radiative Transfer) method.
The overall computational time with this algorithm
scales as $N(\log N)^2$, where $N$ is the number of grid points
in a simulation box.
Therefore, this is a {\it fast} algorithm.
We stress that the computational 
cost with this method is independent of the number
of radiation sources, providing
a natural match to cosmological
simulations where a large number of sources are present.
Local sources and diffuse background are naturally split
and the evolution of each of the two components as well as
interactions between them are 
handled self-consistently.
The devised integral form of the algorithm 
guarantees that the method is completely {\it stable and robust}.

The algorithm is tested on a wide range of analytically tractable problems that 
have significant bearings on cosmological applications.
We find that the algorithm performs very well in all cases,
accurately tracking ionization fronts 
with the error no larger than one cell.
Conservation of photons is observed to be at a level of a few percent.
The accuracy of the results depends weakly on the size
of the time step in all cases;
a time step comparable
to the typical size of a time step for a cosmological hydrodynamic 
simulation is sufficient.
We also apply the algorithm to a realistic density
field produced by a cosmological hydrodynamic simulation
and find that 
conservation of photons
is observed at a level of a few percent.
These extensive tests indicate that 
the algorithm is also very {\it accurate}.

The tests performed is based on a first implementation 
of the basic algorithm outlined in the paper.
It is clear that there is significant room 
for further improvement over the specific implementation.
The current implementation is on a uniform mesh
but we think that variant approaches 
(such as with adaptive mesh refinement, for example)
based on this method are possible and will be explored in the future.

\acknowledgments
The work is supported in part
by grants NAG5-2759, AST91-08103 and ASC93-18185.
This implementation of the algorithm will be fine tuned,
documented and made available in due course.

\end{document}